\documentclass[twocolumn,showpacs,prd,preprintnumbers,amssymb,amsmath,superscriptaddress,floatfix]{revtex4}

\usepackage{amssymb, amsmath}
\usepackage{graphicx}
\usepackage{dcolumn}
\usepackage{bm}
\usepackage[figuresright]{rotating}
\usepackage{fancyhdr}
\usepackage{enumerate}
\usepackage{indentfirst}
\usepackage{xspace}
\usepackage{color}
\usepackage{epsfig}
\usepackage{grffile}
\usepackage{subfigure}

\newcommand{\etal}{{\it et\ al}.}
\newcommand{\pizero}{\pi^{0}}
\newcommand{\pt}{p_\text{T}}

\begin{document}
\title{Measurement of forward neutral pion transverse momentum spectra for
$\sqrt{s}$ = 7TeV proton-proton collisions at the LHC}

\newcommand{\firenzeinfn}{\affiliation{INFN Section of Florence, Italy}}
\newcommand{\firenzeuniv}{\affiliation{University of Florence, Italy}}
\newcommand{\nagoya}{\affiliation{Solar-Terrestrial Environment Laboratory, Nagoya University, Nagoya, Japan}}
\newcommand{\kmi}{\affiliation{Kobayashi-Maskawa Institute for the Origin of Particles and the Universe, Nagoya University, Nagoya, Japan}}
\newcommand{\polyteque}{\affiliation{Ecole-Polytechnique, Palaiseau, France}}
\newcommand{\waseda}{\affiliation{RISE, Waseda University, Japan}}
\newcommand{\cern}{\affiliation{CERN, Switzerland}}
\newcommand{\kanagawa}{\affiliation{Kanagawa University, Japan}}
\newcommand{\cataniainfn}{\affiliation{INFN Section of Catania, Italy}}
\newcommand{\cataniauniv}{\affiliation{University of Catania, Italy}}
\newcommand{\lbn}{\affiliation{LBNL, Berkeley, California, USA}}

\firenzeinfn
\firenzeuniv
\nagoya
\polyteque
\kmi
\waseda
\cataniainfn
\cern
\kanagawa
\cataniauniv
\lbn

%%%%%%%%%%%%%%%%%%%%%%%%%%%%%%%%%%%%%%%
\author{O.~Adriani}
\firenzeinfn
\firenzeuniv

\author{L.~Bonechi}
\firenzeinfn

\author{M.~Bongi}
\firenzeinfn

\author{G.~Castellini}
\firenzeinfn
\firenzeuniv

\author{R.~D'Alessandro}
\firenzeinfn
\firenzeuniv

\author{K.~Fukatsu}
\nagoya

\author{M.~Haguenauer}
\polyteque

\author{T.~Iso}
\nagoya

\author{Y.~Itow}
\nagoya
\kmi

\author{K.~Kasahara}
\waseda

\author{K.~Kawade}
\nagoya

\author{T.~Mase}
\nagoya

\author{K.~Masuda}
\nagoya

%\author{Y.~Matsubara}
%\nagoya

\author{H.~Menjo}
\firenzeinfn
\kmi

\author{G.~Mitsuka}
%\email{mitsuka@stelab.nagoya-u.ac.jp}
\nagoya

\author{Y.~Muraki}
\nagoya

\author{K.~Noda}
\cataniainfn

\author{P.~Papini}
\firenzeinfn

\author{A.-L.~Perrot}
\cern

\author{S.~Ricciarini}
\firenzeinfn

\author{T.~Sako}
\nagoya
\kmi

\author{Y.~Shimizu}
\waseda

\author{K.~Suzuki}
\nagoya

\author{T.~Suzuki}
\waseda

\author{K.~Taki}
\nagoya

\author{T.~Tamura}
\kanagawa

\author{S.~Torii}
\waseda

\author{A.~Tricomi}
\cataniainfn
\cataniauniv

\author{W.~C.~Turner}
\lbn

\collaboration{The LHCf Collaboration}
\noaffiliation

\date{\today}

% ---- Abstract ----
\begin{abstract}

The inclusive production rate of neutral pions in the rapidity range greater
than $y=8.9$ has been measured by the Large Hadron Collider forward (LHCf)
experiment during $\sqrt{s}=7$\,TeV proton-proton collision operation in early
2010. This paper presents the transverse momentum spectra of the neutral pions.
The spectra from two independent LHCf detectors are consistent with each other
and serve as a cross-check of the data. The transverse momentum spectra are also
compared with the predictions of several hadronic interaction models that are
often used for high-energy particle physics and for modeling ultra-high-energy
cosmic-ray showers.

\end{abstract}

% --- PACS number ---

\maketitle

% ----- Introduction -----
\section{Introduction}\label{sec:introduction}

One of the important tasks of strong-interaction physics described by Quantum
Chromodynamics (QCD) is to provide a detailed understanding of forward particle
production in hadronic interactions. QCD involves two types of limiting
processes: ``hard'' and ``soft''.

Hard processes occur in the range characterized by a large four-momentum
transfer $t$, where $|t|$ should be larger than 1\,GeV$^2$. Note that units used
in this report are $c = k$ (Boltzmann constant) $ = 1$. Deep inelastic
scattering that is accompanied by the exchange of virtual photons or vector
bosons, or jets produced by large transverse momentum ($\pt$) partons are
typical phenomena that are categorized as hard processes. The hard processes
have been successfully described by perturbation theory, owing to the asymptotic
freedom of QCD at high energy.

On the other hand, soft processes occur when the four-momentum transfer $|t|$ is
smaller than 1\,GeV$^2$. These processes, which correspond to a large impact
parameter, have a large QCD coupling constant and cannot be calculated by
perturbative QCD. Gribov-Regge theory is applicable for describing soft
processes~\cite{Gribov, Regge} and the Pomeron contribution, as a component of
the Gribov-Regge approach to high-energy hadronic interactions, increases with
increasing energy~\cite{Pomeron} and should dominate at the TeV energy scale.
However there still exists a problem for the theories that involve these virtual
quasi-particles.
Since the treatment of the Pomeron differs amongst the model theories they
predict different results for particle production.
Thus a deeper understanding of soft processes is needed and soft processes are
mostly equivalent to forward or large rapidity particle production in hadronic
interactions. However experimental data for large rapidity are meager.
Moreover the experimental data that do exist have so far been carried out at
relatively low energy, for example ISR~\cite{ISR} at $\sqrt{s} =$ 53\,GeV and
UA7~\cite{UA7} at $\sqrt{s} =$ 630\,GeV.

The Large Hadron Collider forward (LHCf) experiment~\cite{LHCfTDR} has been
designed to measure the hadronic production cross sections of neutral particles
emitted in very forward angles in proton-proton collisions at the LHC, including
zero degrees. The LHCf detectors have the capability for precise measurements of
forward high-energy inclusive-particle-production cross sections of photons,
neutrons, and possibly other neutral mesons and baryons.
Among the many secondary neutral particles that LHCf can detect, the $\pizero$
mesons are the most sensitive to the details of the proton-proton interactions.
Thus a high priority has been given to analyzing forward $\pizero$ production
data in order to provide key information for an as yet un-established hadronic
interaction theory at the TeV energy scale.
The analysis in this paper concentrates on obtaining the inclusive production
rate for $\pizero$s in the rapidity range larger than $y=8.9$ as a function of
the $\pizero$ transverse momentum.
 
In addition to the aim described above, this work is also motivated by an
application to the understanding of Ultra-High-Energy Cosmic-Ray (UHECR)
phenomena, which are sensitive to the details of soft $\pizero$ production at
extreme energy. It is known that the lack of knowledge about forward particle
production in hadronic collisions hinders the interpretation of observations of
UHECR~\cite{Ulrich, Parsonsa}. Although UHECR observations have made notable
advances in the last few years~\cite{AugerSpectrum, AugerAniso, AugerComp,
HiResSpectrum, HiResAniso, HiResComp, TA}, critical parts of the analysis depend
on Monte Carlo (MC) simulations of air shower development that are sensitive to
the choice of the hadronic interaction model. It should also be remarked that
the LHC has reached 7\,TeV collision energy, which in the laboratory frame of
UHECR observations is equivalent to $2.6\times10^{16}$\,eV, and this
energy is above the ``knee'' region of the primary cosmic ray energy spectrum ($\sim
4\times10^{15}$\,eV)~\cite{Horandel}. The data provided by LHCf should then
provide a useful bench mark for the MC codes that are used for the simulation of
UHECR atmospheric showers.

This paper is organized as follows. In Sec.~\ref{sec:detector} the LHCf
detectors are described. Sec.~\ref{sec:data} summarizes the conditions for
taking data and the MC simulation methodology. In Sec.~\ref{sec:framework} the
analysis framework is described. The factors that contribute to the systematic
uncertainty of the results are explained in Sec.~\ref{sec:systerror} and the
analysis results are then presented in Sec.~\ref{sec:result}.
Sec.~\ref{sec:discussions} discusses the results that have been obtained and
compare these with the predictions of several hadronic interaction models.
Finally, concluding remarks are found in Sec.~\ref{sec:conclusions}.

%
% ----- Detector -----
%
\section{The LHCf detectors}\label{sec:detector}

\begin{figure}[htbp]
  \begin{center}
  \includegraphics[width=4.2cm, keepaspectratio]{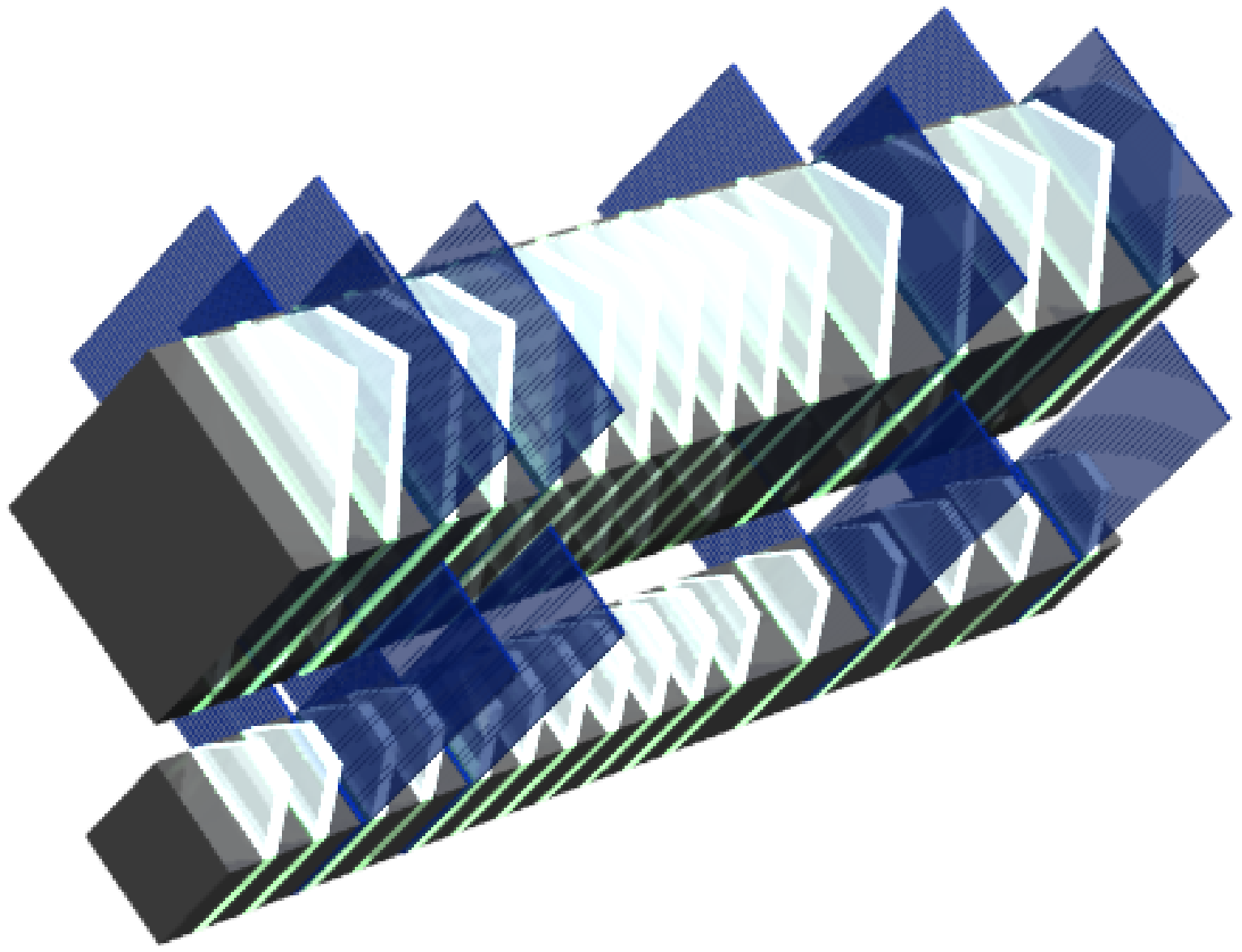}
  \includegraphics[width=4.2cm, keepaspectratio]{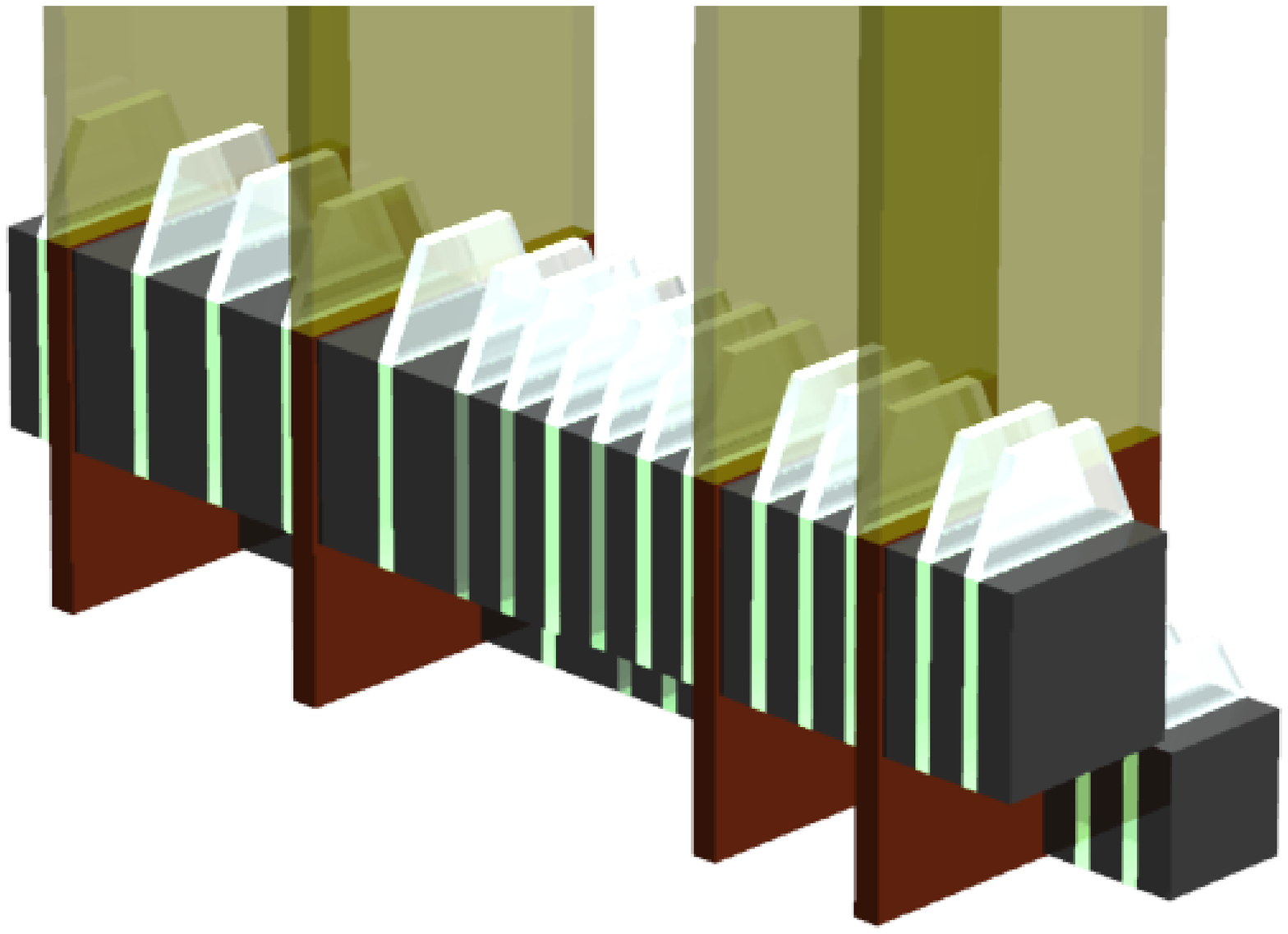}
  \caption{(color online). Schematic views of the Arm1 (left) and Arm2 (right)
  detectors. The transverse sizes of the calorimeters are 20\,$\times$20\,mm$^2$
  and 40\,$\times$40\,mm$^2$ in Arm1, and 25\,$\times$25\,mm$^2$ and
  32\,$\times$32\,mm$^2$ in Arm2.}
  \label{fig:detectors}
  \end{center}
\end{figure}

Two independent LHCf detectors, called Arm1 and Arm2, have been installed in the
instrumentation slots of the target neutral absorbers (TANs)~\cite{TAN} located
$\pm$140\,m from the ATLAS interaction point (IP1) and at zero degree collision
angle. Fig.~\ref{fig:detectors} shows schematic views of the Arm1 (left) and
Arm2 (right) detectors. Inside a TAN the beam-vacuum-chamber makes a Y-shaped
transition from a single common beam tube facing IP1 to two separate beam tubes
joining to the arcs of the LHC.
Charged particles produced at IP1 and directed towards the TAN are swept aside
by the inner beam separation dipole magnet D1 before reaching the TAN.
Consequently only neutral particles produced at IP1 enter the LHCf detector.
At this location the LHCf detectors cover the pseudorapidity range from 8.7 to
infinity for zero degree beam crossing angle. With a maximum beam crossing angle
of 140\,$\mu$rad, the pseudorapidity range can be extended to 8.4 to infinity.

Each LHCf detector has two sampling and imaging calorimeters composed of 44
radiation lengths ($X_0$) of tungsten and 16 sampling layers of 3\,mm thick
plastic scintillator. The transverse sizes of the calorimeters are
20\,$\times$20\,mm$^2$ and 40\,$\times$40\,mm$^2$ in Arm1, and
25\,$\times$25\,mm$^2$ and 32\,$\times$32\,mm$^2$ in Arm2. The smaller
calorimeters cover zero degree collision angle. Four X-Y layers of position
sensitive detectors are interleaved with the layers of tungsten and scintillator
in order to provide the transverse positions of the showers.
Scintillating fiber (SciFi) belts are used for the Arm1 position sensitive
layers and silicon micro-strip sensors are used for Arm2. Readout pitches are
1\,mm and 0.16\,mm for Arm1 and Arm2, respectively.

More detail on the scientific goals and the construction and performance of the
detectors can be found in previous reports~\cite{LHCfJINST, prototype, SPS2007,
LHCfsilicon, menjopi0}.

%
% ----- Data summary -----
%
\section{Summary of the conditions for taking data and of the methodology for
performing Monte Carlo simulations}\label{sec:data}

\subsection{Conditions for taking experimental data}\label{sec:measured_data}

The experimental data used for the analysis of this paper were obtained on May
15th and 16th 2010 during proton-proton collisions at $\sqrt{s}$=7\,TeV with
zero degree beam crossing angle (LHC Fill 1104). Data taking was carried out in
two different runs: the first run was on May 15th from 17:45 to 21:23, and the
second run was on May 16th from 00:47 to 14:05. The events that were recorded
during a luminosity optimization scan and a calibration run were removed from
the data set for this analysis.

The range of total luminosity of the three crossing bunch pairs was ${\cal L} =
(6.3-6.5) \times 10^{28} $cm$^{-2}$s$^{-1}$ for the first run and ${\cal L} =
(4.8-5.9) \times 10^{28} $cm$^{-2}$s$^{-1}$ for the second run. These ranges of
luminosity were ideal for the LHCf data acquisition system.
The integrated luminosities for the data analysis reported in this paper were
derived from the counting rate of the LHCf Front Counters~\cite{LHCfLuminosity},
and were 2.53\,nb$^{-1}$ (Arm1) and 1.90\,nb$^{-1}$ (Arm2) after taking the live
time percentages into account. The average live time percentages for the
first/second run were 85.7\,\%/81.1\,\% for Arm1 and 67.0\,\%/59.7\,\% for Arm2.
The live time percentages for the second run were smaller than the first run
owing to a difference in the trigger schemes. In both runs the trigger
efficiency achieved was $>$99\% for photons with energy $E >
100$\,GeV~\cite{photonpaper}.

The events containing more than one collision in a single bunch crossing
(pile-up events) could potentially cause a bias in the $\pt$ spectra.
For example combinatorial single-hits from different collisions within a single
bunch crossing might be identified as multi-hit events from a single collision
and removed from the analysis. (Multi-hit events have two showers in a single
calorimeter and are eliminated from the data analysis. The production rates are
later corrected for this cut. See Fig.~\ref{fig:pi0kinematics} and related
discussion.) However it can be shown that pile-up events are negligible for the
LHCf data taking conditions of this report.
Given that a collision has occurred, the probability of pile-up
($P_\text{pileup}$) is calculated from the Poisson probability distribution for
$n$ collisions $P_\text{poi}(n)$ according to $P_\text{pileup} =
P_\text{poi}(n\ge2) / P_\text{poi}(n\ge1)$.
With the highest bunch luminosity ${\cal L} = 2.3 \times 10^{28}
$cm$^{-2}$s$^{-1}$ used in this analysis, an inelastic cross section
$\sigma_\text{inel}$ = 73.6\,mb and the revolution frequency of LHC
$f_\text{rev}$ = 11.2\,kHz, the pile-up probability is $P_\text{pileup}
\sim0.07$.
However considering that the acceptance of the LHCf calorimeter for inelastic
collisions is $\sim$0.03, only 0.2\% of events have more than one shower event
in a single calorimeter due to pile-up and this is negligible.
%% Turner comment, #1
% they can be eliminated in the multi-hit cut used for event selection. In this
% analysis, the detection efficiency for the multi-hit events is greater than 70\%
% for Arm1 and 90\% for Arm2 as discussed later in
% Sec.~\ref{sec:positionreconstruction}.
% Thus pile-up events remained in the data set are estimated to be smaller than
% $10^{-3}$ relative to the single-hit events, and have no significant effect on
% the results of this report.

Detailed discussions of pile-up effects and background events from collisions
between the beam and residual gas molecules in the beam tube can be found in
previous reports~\cite{LHCfTDR, photonpaper}.

\subsection{Methodology for performing Monte Carlo simulations}
\label{sec:mc_simulation}

MC simulation consists of three steps: (1) proton-proton interaction event
generation at IP1, (2) transport from IP1 to the LHCf detectors and (3) the
response of the LHCf detectors.

Proton-proton interaction events at $\sqrt{s} = 7$\,TeV and the resulting flux
of secondary particles and their kinematics are simulated with {\sc cosmos}
(version 8.81). {\sc cosmos} acts as the front end for the external hadronic
interaction models ({\sc qgsjet} II-03~\cite{QGS2}, {\sc dpmjet
3.04}~\cite{DPM3}, {\sc sibyll 2.1}~\cite{SIBYLL} and {\sc epos
1.99}~\cite{EPOS}) that describe the proton-proton interactions. While {\sc
pythia 8.145}~\cite{PYTHIA8a, PYTHIA8b} serves as its own front end for the
generation of proton-proton interaction events.

Next, the generated secondary particles are transported in the beam pipe from
IP1 to the TAN, taking account of the deflection of charged particles by the Q1
quadrupole and D1 beam separation dipole, particle decay, and particle
interaction with the beam pipe and the Y-shaped beam-vacuum-chamber transition
made of copper (1\,$X_0$ projected thickness in front of the LHCf detectors).
Charged particles are swept away by the D1 magnet before reaching the LHCf
detectors. This simulation uses the {\sc epics} library~\cite{EPICS} (version
7.49) and a part of {\sc cosmos}.
{\sc epics} deals with the transport of secondary particles. Particle
interactions with the residual gas molecules inside the beam pipe are not
simulated. Contamination from beam-gas background events in the data set used
for analysis is estimated to be only $\sim0.1$\,\% and has no significant impact
on the $\pt$ spectra reported.

Finally the simulations of the showers produced in the LHCf detectors and their
response are carried out for the particles arriving at the TAN using the {\sc
cosmos} and {\sc epics} libraries.
The survey data for detector position and random fluctuations equivalent to
electrical noise are taken into account in this step.
The Landau-Pomeranchuk-Migdal effect~\cite{LPMa, LPMb} that longitudinally
lengthens an electromagnetic shower at high energy is also considered. A change
of the $\pt$ spectra caused by LPM effects is only at the 1\% level since the
reconstruction of energy deposited in the calorimeters is carried out to a
sufficiently deep layer whereby the energy of electromagnetic showers is almost
perfectly deposited within the calorimeter.

The simulations of the LHCf detectors are tuned to test beam data taken at the
CERN SPS in 2007~\cite{SPS2007}. The validity of the detector simulation was
checked by comparing the shower development and deposited energy for each
calorimeter layer to the results obtained by the {\sc fluka}
library~\cite{FLUKA}.

In order to validate the reconstruction algorithms and to estimate a possible
reconstruction bias beyond the energy range of the SPS test beam results, the MC
simulations are generated for $1.0\times10^{8}$ inelastic collisions, where the
secondary particles are generated by the {\sc epos} 1.99~\cite{EPOS} hadronic
interaction model. This MC simulation is referred to as the ``reference MC
simulation'' in the following text.

Similarly the ``toy MC simulations'' discussed below are performed in order to
determine various correction factors to use in the event reconstruction
processes. In the toy MC simulations, a single-photon with a given fixed energy
is directed at the LHCf detectors.

%
% ----- Analysis framework -----
%
\section{Analysis framework}\label{sec:framework}

% ----- Event reconstruction -----
\subsection{Event reconstruction and selection}\label{sec:reconstruction}

\begin{figure}[htbp]
  \begin{center}
  \includegraphics[width=4.1cm, keepaspectratio]{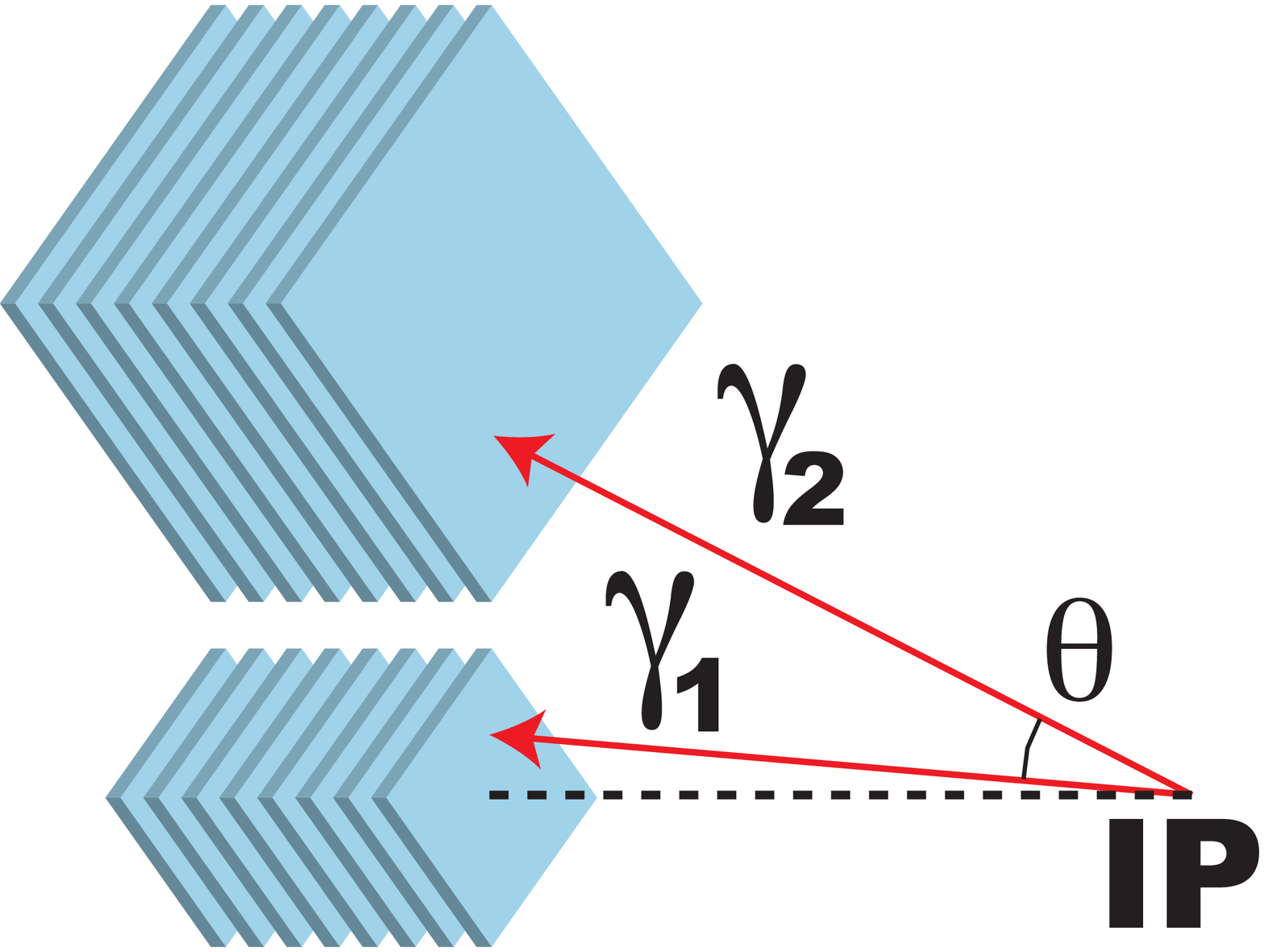}
  \includegraphics[width=4.1cm, keepaspectratio]{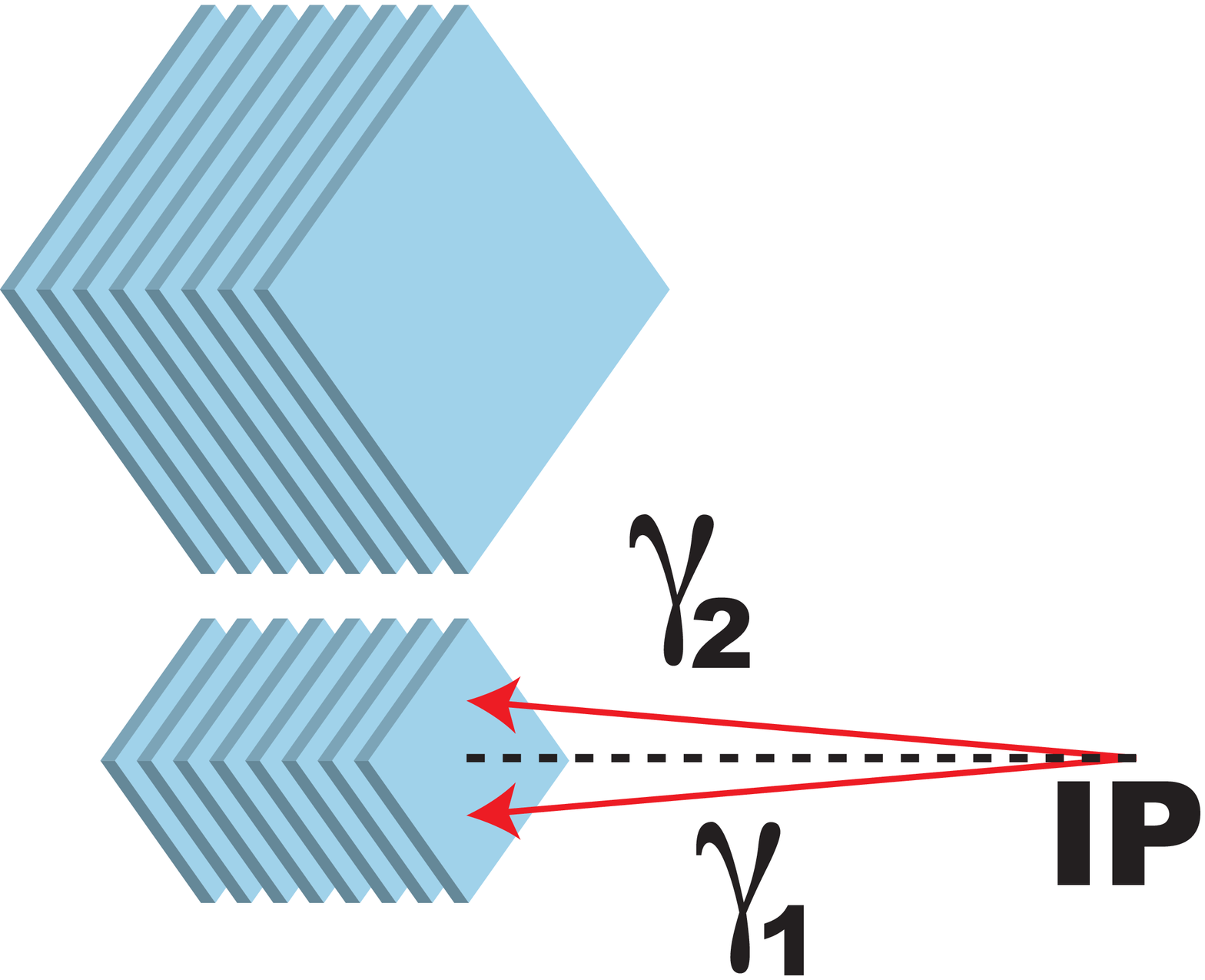}
  \caption{(color online). Observation of $\pizero$ decay by a LHCf detector.
  (Left) Two photons enter different calorimeters. (Right) Two photons enter one
  calorimeter.}
  \label{fig:pi0kinematics}
  \end{center}
\end{figure}

Observation of $\pizero$ mesons by a LHCf detector is illustrated in
Fig.~\ref{fig:pi0kinematics}. The $\pizero$s are identified by their decay into
two photons. Since the $\pizero$s decay very close to their point of creation at
IP1, the opening angle ($\theta$) between the two photons is the transverse
distance between photon impact points at the LHCf detectors divided by the
distance from IP1 ($z=\pm$141.05\,m).
Consequently the opening angle for the photons from $\pizero$ decay that are
detected by a LHCf detector is constrained by $\theta \lesssim 0.4$\,mrad for
Arm1 and $\theta \lesssim 0.6$\,mrad for Arm2.
Other kinematic variables of the $\pizero$s (energy, $\pt$, and rapidity) are
also reconstructed by using the photon energy and incident position measured by
each calorimeter. Note that for the analysis of this paper events having two
photons entering the same calorimeter (multi-hit events) are rejected (right
panel of Fig.~\ref{fig:pi0kinematics}). The accuracy of energy reconstruction
for such events is still under investigation.
The final inclusive production rates reported in this paper are corrected for
this cut.
In order to ensure good event reconstruction efficiency, the range of the
$\pizero$ rapidity and $\pt$ are limited to $8.9 < y < 11.0$ and $\pt <
0.6$\,GeV, respectively.
% All other particles emitted into the outside of the aperture of the LHCf
% detectors are ignored in this analysis, thus the reported production rates are
% inclusive.
All particles other than photons from $\pizero$ decay are ignored in this
analysis. Thus, also according to the multi-hit $\pizero$ correction described
in detail in Sec.~\ref{sec:multihitpizero}, the reported production rates are
inclusive.
The standard reconstruction algorithms are described in this section and
systematic uncertainties will be discussed in Sec.~\ref{sec:systerror}.

% ----- Hit position reconstruction -----
\subsubsection{Hit position reconstruction}\label{sec:positionreconstruction}

The transverse impact positions of particles entering the calorimeters are
determined using the information provided by the position sensitive layers.
In this analysis, the transverse impact position of the core of an
electromagnetic shower is taken from the position of the peak signal on the
position sensitive layer that has the largest energy deposited amongst all the
position sensitive layers.
% Non-uniformity of light-yield collection by the layers of plastic
% scintillator~\cite{prototype} depends on the incident position of a particle
% hit and is corrected according to the reconstructed position information.

Hit positions that fall within 2\,mm of the edges of the calorimeters are
removed from analysis due to the large uncertainty in the energy determination
of such events owing to shower leakage. For the toy MC simulations, the position
reconstruction resolution is defined as the one standard deviation difference
between the true primary photon position and the reconstructed position of the
shower axis.
% Thus 68\,\% of events have the difference between true photon
% position and reconstructed shower position within plus or minus the resolution
% so defined.
The estimated resolution using the toy MC simulations and test beam data for a
single photon with energy $ E > 100$\,GeV is better than 200\,$\mu$m and
100\,$\mu$m for Arm1 and Arm2, respectively~\cite{LHCfsilicon, Scifi}.

Multi-hit events defined to have more than one photon registered in a single
calorimeter are eliminated from the analysis in this paper. Multi-hit candidates
that have two distinct peaks in the lateral shower impact distribution are
searched for using the algorithm that has been implemented in the
TSpectrum~\cite{TSpectrum} class in {\sc root}~\cite{ROOT}. When the separation
between peaks is greater than 1\,mm and the lower energy photon has more than
5\,\% of the energy of the nearby photon, the MC simulation estimated
efficiencies for identifying multi-hit events are larger than 70\,\% and 90\,\%
for Arm1 and Arm2, respectively~\cite{photonpaper}.
The efficiency for Arm2 is better than that for Arm1 owing to the finer readout
pitches of the silicon micro-strip sensors. The subtraction of the remaining
contamination by multi-hit events is discussed in Sec.~\ref{sec:bgsubtraction}.

On the other hand for single-hit events not having two identifiable peaks, the
MC simulation estimated efficiency for correctly identifying true single photon
events with energy $E > 100$\,GeV is better than 98\,\% both for Arm1 and Arm2,
although the precise percentage depends slightly on the photon energy.

% ----- Energy reconstruction -----
\subsubsection{Energy reconstruction}\label{sec:energyreconstruction}

% The energy of a primary photon incident on a calorimeter $E$[GeV] is calculated
% according to
% \begin{equation}
% 	E = A_E {dE}^2 + B_E dE + C_E.
% 	\label{eq:bgsubtraction1}
% \end{equation}
% where $dE$ is the number of minimum ionizing shower particles (MIPs) summed over
% the 2nd to 13th scintillating layers and the coefficients $A_E$[GeV/MIP$^2$],
% $B_E$[GeV/MIP] and $C_E$[GeV] have been determined from the response of the
% calorimeters to SPS electron test beam data taken below 200 GeV, muon test beam
% data taken at 150 GeV~\cite{prototype} and toy Monte Carlo calculations
% above 200 GeV. The test beam electrons of known energy simulate an incident primary
% photon and the test beam muons calibrate the response of the electronics to a
% minimum ionizing particle, defined here to be the most probable ionization
% produced by a 150 GeV muon. The most probable energy deposited by a 150 GeV muon
% passing through a 3 mm thick plastic scintillator is 0.453 MeV. The energy
% resolution for single photons above 100 GeV is $\sigma(E)/E =
% 36.5(37.1)\%/\sqrt{E(\text{GeV})} + 1.1(0.4)\%$ for the 20mm$^2$ and 40mm$^2$
% Arm1 calorimeters.
The charge information in each scintillation layer is converted to a deposited
energy by using calibration factors obtained from the SPS electron test beam
data taken below 200\,GeV~\cite{prototype}.
In this analysis the deposited energy is scaled to the number of minimum
ionizing shower particles with a coefficient 1\,MIP = 0.453\,MeV that
corresponds to the most probable deposited energy by a 150\,GeV muon passing
through a 3\,mm thick plastic scintillator. The sum of the energy deposited in
the 2$^{nd}$ to 13$^{th}$ scintillation layers ($dE$\,[MIP]) is then converted
to the primary photon energy $E$[GeV] using a polynomial function
\begin{equation}
	E = A_E {dE}^2 + B_E dE + C_E.
	\label{eq:miptoenergy}
\end{equation}
The coefficients $A_E$\,[GeV/MIP$^2$], $B_E$\,[GeV/MIP] and $C_E$\,[GeV] are
determined from the response of the calorimeters to single photons by the toy MC
simulations. The validity of this method has been confirmed with the SPS beam
tests. The MC estimated energy resolution for single photons above 100\,GeV
considering the LHC data taking situation is given by the expression
\begin{equation}
%	\sigma(E)/E = \sigma_\text{res}/\sqrt{E/100\,\text{GeV}} \oplus C_\text{res}.
	\sigma(E)/E \sim 8\,\%/\sqrt{E/100\,\text{GeV}} \oplus 1\,\%.
	\label{eq:energyres}
\end{equation}
% The coefficients $\sigma_\text{res}$ and $C_\text{res}$ are summarized in
% Table~\ref{tab:energyres}.
% 
% \begin{table}
%   \begin{center}
%     \begin{tabular}{c|c|c|c|c}
%       \hline
%        & Arm1 20\,mm & Arm1 40\,mm & Arm2 25\,mm & Arm2 32\,mm \\
%       \hline
%        $\sigma_\text{res}$ & 7.4\,\% & 7.4\,\% & 8.5\,\% & 7.3\,\% \\
%        $C_\text{res}$      & 1.1\,\% & 0.5\,\% & 1.0\,\% & 1.0\,\% \\
%       \hline
%     \end{tabular}
%     \caption{Summary of energy resolution for single photons above 100\,GeV.}
%     \label{tab:energyres}
%   \end{center}
% \end{table}
 
Corrections for shower leakage effects~\cite{prototype, LHCfJINST} are carried
out during the energy reconstruction process. Corrections are applied for
leakage of particles out of the calorimeters and for leakage of particles in
that have escaped from the adjacent calorimeter. Both of the leakage effects
depend on the transverse location of the shower axis in the calorimeters. The
correction factors have been estimated from the toy MC simulations.
The light-yield collection efficiency of the plastic scintillation
layers~\cite{prototype} is also a function of the transverse location of the
shower axis and corrected for in this step.

Events having a reconstructed energy below 100\,GeV are eliminated from the
analysis firstly to reject particles produced by interaction of collision
products with the beam pipe, and secondly to avoid errors due to trigger
inefficiency (see Sec.~\ref{sec:measured_data}).
% The quadratic term described by $A_E$ in Eq.~\ref{eq:bgsubtraction1} is only
% important below 100\,GeV and is therefore not important for the analysis of this
% paper.

% ----- Particle identification -----
\subsubsection{Particle identification}\label{sec:pid}

\begin{figure}[htbp]
  \begin{center}
  \includegraphics[width=6cm, keepaspectratio]{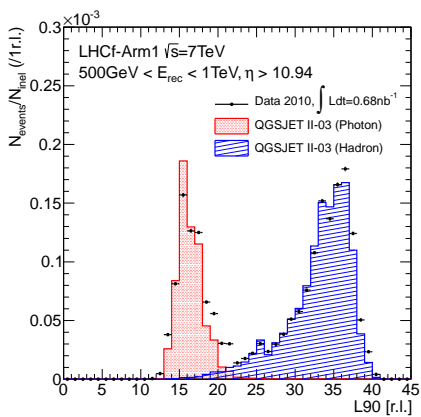}
  \caption{(color online). $L_{90\%}$ distribution measured by the Arm1-20$\,$mm
  calorimeter for the reconstructed energy of 500\,GeV--1\,TeV.}
  \label{fig:L90}
  \end{center}
\end{figure}

The particle identification (PID) process is applied in order to select pure
electromagnetic showers, specifically photons from $\pizero$ decay, and to
reduce hadron contamination, specifically from neutrons.
A parameter $L_{90\%}$ is defined for this purpose. $L_{90\%}$ is the
longitudinal distance, in units of radiation length, measured from the 1st
tungsten layer of a calorimeter to the position where the energy deposition
integral reaches 90\,\% of the total shower energy deposition.
Fig.~\ref{fig:L90} shows the distribution of $L_{90\%}$ for the 20\,mm
calorimeter of the Arm1 detector for events having a reconstructed energy in the
range 500\,GeV$<E<$1\,TeV. Experimental data (black dots) and the MC simulations
based on {\sc qgsjet} II-03 (shaded areas) are shown. The normalization factors
of pure photon and pure hadron incident events are modified to get the best
agreement between the $L_{90\%}$ distributions of the experimental data and the
MC simulations.
The best agreement is obtained by a chi-square test of the $L_{90\%}$
distribution of the experimental data relative to the MC simulation. The two
distinct peaks correspond to photon ($L_{90\%} \lesssim 20\,X_0$) and hadron
($L_{90\%} \gtrsim 20\,X_0$) events.

PID criteria that depend on the energy of the individual photons are defined in
terms of the $L_{90\%}$ distribution in order to keep the $\pizero$ selection
efficiency at approximately 90\,\% over the entire $\pt$ range.
These criteria $f_{L90\%}(E_1, E_2)$ are expressed as a function of the photon
energies measured by the small ($E_1$) and large ($E_2$) calorimeters and have
been determined by the toy MC simulations for each Arm.
The remaining hadron contamination is removed by background subtraction
introduced in Sec.~\ref{sec:bgsubtraction}.
The unavoidable selection inefficiency of 10\,\% is corrected for in the
unfolding process to be discussed later (Sec.~\ref{sec:unfolding}).

Table~\ref{tbl:eventselection} summarizes the $\pizero$ event selection
criteria that are applied prior to reconstruction of the $\pizero$ kinematics.

\begin{table}
  \begin{center}
    \begin{tabular}{c|c}
      \hline
      Incident position & within 2\,mm from the edge of calorimeter \\
      \hline
      Energy threshold  & $E_\text{photon} > 100$\,GeV \\
      \hline
      Number of hits    & Single-hit in each calorimeter \\
      \hline
      PID & Photon like ($L_{90\%} < f_{L90\%}(E_1, E_2)$) \\
      \hline
    \end{tabular}
    \caption{Summary of criteria for event selections of the $\pizero$ sample.}
    \label{tbl:eventselection}
  \end{center}
\end{table}

% ----- Pizero reconstruction -----
\subsection{$\pizero$ reconstruction}\label{sec:pi0reconstruction}

Candidates for $\pizero$ events are selected using the characteristic peak in
the two-photon invariant mass distribution corresponding to the $\pizero$ rest
mass. Reconstruction of the invariant mass $m_{\gamma\gamma}$ is done using the
incident positions and energies information of the photon pair,
\begin{equation}
	m^2_{\gamma\gamma} = (q_1 + q_2)^2 \thickapprox E_1E_2\theta^2, 
	\label{eq:pi0mass}
\end{equation}
where $q_i$ and $E_i$ are the energy-momentum 4-vectors and energies of the
decay photons in the laboratory frame, respectively. $\theta$ is the opening
angle between the two photons in the laboratory frame. The last approximation in
Eq.~(\ref{eq:pi0mass}) is valid since the $\pizero$s decay very close to IP1
(mean $\pizero$ flight path $\lesssim$ 1\,mm).
This approximation and the reconstruction algorithm for $\pizero$ events have
been verified by analysis of the reference MC simulations of the energy,
rapidity and $\pt$ of the $\pizero$s. The reconstructed invariant mass is
concentrated near peaks at 135.2$\pm$0.2\,MeV in Arm1 and 134.8$\pm$0.2\,MeV in
Arm2, thus reproducing the $\pi^{0}$ mass. The uncertainties given for the mass
peaks are statistical only.

It should be noted however that in the $\pizero$ analysis of the experimental
LHCf data energy scale corrections are needed so the $\pizero$ mass peaks for
Arm1 and Arm2 occur at the proper value. With no energy scale corrections
applied to the LHCf data, the reconstructed invariant mass peaks using gain
calibration constants determined by test beam data occur at 145.8$\pm$0.1\,MeV
(Arm1) and 139.9$\pm$0.1\,MeV (Arm2).
Therefore energy scale corrections of $-8.1$\,\% (Arm1) and $-3.8$\,\% (Arm2)
applied to the raw measured photon energies are needed to bring the
reconstructed $\pizero$ rest mass into agreement with the world averaged
$\pizero$ rest mass~\cite{PDG}. The cause of these energy scale corrections is
probably due to a temperature dependent shift of PMT gain.
However at this point the temperature dependent shift of PMT gain is only
qualitatively understood.
%% Turner comment
% However at this point this cause is only qualitatively understood and degeneracy
% of the gain change in terms of the mass shift between PMTs of the small and
% large calorimeter is not solved.
Note that the typical uncertainty in opening angle is estimated to be less than
1\,\% relative to the reconstructed invariant mass by the position determination
resolution and the alignment of the position sensitive detectors.

% ----- Background subtraction -----
\subsection{Background subtraction}\label{sec:bgsubtraction}

Background contamination of two-photon $\pizero$ events by hadron events and the
accidental coincidence of two photons not coming from the decay of a single
$\pizero$ are subtracted using the so called ``sideband'' method.

\begin{figure}[htbp]
  \begin{center}
  \includegraphics[width=6cm, keepaspectratio]{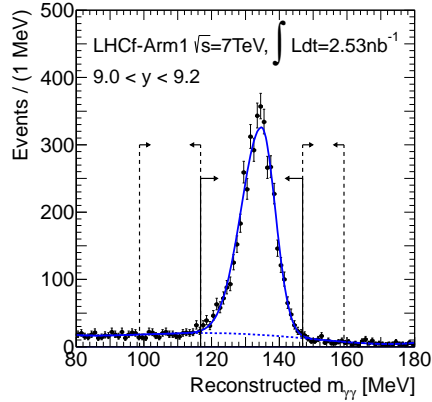}
  \caption{(color online). Reconstructed invariant mass distribution within the
  rapidity range from 9.0 to 9.2. Solid curve shows the best-fit composite
  physics model to the invariant mass distribution. Dashed curve indicates the
  background component. Solid and dashed curves indicate the signal and
  background windows, respectively.}
  \label{fig:mass-fit}
  \end{center}
\end{figure}

Fig.~\ref{fig:mass-fit} shows an example of the reconstructed two-photon
invariant mass distribution of the experimental data of Arm1 in the rapidity
range from 9.0 to 9.2.
The energy scale correction discussed in the previous section has been applied.
The sharp peak around 135\,MeV is due to $\pizero$ events. The solid curve
represents the best-fit of a composite physics model to the invariant mass
distribution of the data.
The model consists of an asymmetric Gaussian distribution (also known as a
bifurcated Gaussian distribution) for the signal component and a 3rd order
Chebyshev polynomial function for the background component. The dashed curve
indicates the background component.

Using the expected mean ($\hat{m}$) and 1\,$\sigma$ deviations ($\sigma_l$ for
lower side and $\sigma_u$ for upper side) of the signal component, the signal
window is defined as the invariant mass region within the two solid arrows shown
in Fig.~\ref{fig:mass-fit}, where the lower and upper limits are given by
$\hat{m}-3\sigma_l$ and $\hat{m}+3\sigma_u$, respectively.
The background window is constructed from the two sideband regions,
[$\hat{m}-6\sigma_l$, $\hat{m}-3\sigma_l$] and [$\hat{m}+3\sigma_u$,
$\hat{m}+6\sigma_u$], that are defined as the invariant mass regions within the
dashed arrows in Fig.~\ref{fig:mass-fit}.

The rapidity and $\pt$ distributions of the signal ($f(y, \pt)^\text{Sig}$) are
then obtained by subtracting the background distribution ($f(y,
\pt)^\text{BG}$), estimated by the background window, from the signal-rich
distribution ($f(y, \pt)^\text{Sig+BG}$) selected from the signal window. The
fraction of the background component included in the signal window can be
estimated using the likelihood function ($L_\text{BG}(y, \pt,
m_{\gamma\gamma})$) characterized by the best-fit 3rd order Chebyshev polynomial
function. For simplicity, $L_\text{BG}(y, \pt, m_{\gamma\gamma})$ is shortened
as $L_\text{BG}$ in the following text. Thus the signal distribution with
background subtracted is given by
\begin{eqnarray}
	f(y, \pt)^\text{Sig} &=& f(y, \pt)^\text{Sig+BG} - \\ \nonumber
	&~& R(y, \pt, \hat{m}, \sigma_l, \sigma_u)f(y, \pt)^\text{BG},
	\label{eq:bgsubtraction1}
\end{eqnarray}
where $R(y, \pt, \hat{m}, \sigma_l, \sigma_u)$ is the
normalization for the background distribution and written as
\begin{eqnarray}
	&&\hspace{-7mm}R(y, \pt, \hat{m}, \sigma_l, \sigma_u) = \\ \nonumber
	&&\hspace{-6mm} \frac{
 	\int_{\hat{m}-3\sigma_l}^{\hat{m}+3\sigma_u}L_\text{BG}dm_{\gamma\gamma}
 	}{
 	\int_{\hat{m}-6\sigma_l}^{\hat{m}-3\sigma_l}L_\text{BG}dm_{\gamma\gamma}
 	+
 	\int_{\hat{m}+3\sigma_u}^{\hat{m}+6\sigma_u}L_\text{BG}dm_{\gamma\gamma}
 	}.
	\label{eq:bgsubtraction2}
\end{eqnarray}

% ----- Unfolding -----
\subsection{Unfolding of spectra}\label{sec:unfolding}

The raw rapidity -- $\pt$ distributions must be corrected for unavoidable
reconstruction inefficiency and for the smearing caused by finite position and
energy resolutions.
An iterative Bayesian method~\cite{dagostini-1, dagostini-2} is used to
simultaneously correct for both effects. The advantages of an iterative Bayesian
method with respect to other unfolding algorithms are discussed in another
report~\cite{dagostini-1}. The unfolding procedure for the data is organized as
follows.

First, the response of the LHCf detectors to single $\pizero$ events is
simulated by toy MC calculations. In the toy MC simulations, two photons
from the decay of $\pizero$s and low energy background particles such as those
originating in a prompt photon event or a beam-pipe interaction are traced
through the detector and then reconstructed with the event reconstruction
algorithm introduced above.
Note that the single $\pizero$ kinematics that are simulated within the allowed
phase space are independent of the particular interaction model that is being
used.
The background particles are simulated by a hadronic interaction model which is
discussed later, since the amount of background particles is not directly
measured by the LHCf detector.

The detector response to $\pizero$ events depends on rapidity and $\pt$, since
the performance of the particle identification algorithm and the selection
efficiency of events with a single photon hit in both calorimeters depend upon
the energy and the incident position of a particle.
The reconstructed rapidity -- $\pt$ distributions for given true rapidity --
$\pt$ distributions then lead to the calculation of the response function. Then
the reconstructed rapidity and $\pt$ spectra are corrected with the response
function which is equivalent to the likelihood function in Bayes' theorem. The
corrections are carried out iteratively whereby the starting point of the
current iteration is the ending point of the previous iteration.
Statistical uncertainty is also propagated from the first iteration to the last.
Iteration is stopped at or before the 4th iteration to obtain a regularization
of the unfolded events.

Validation of the unfolding procedure is checked by applying the response
function to the reference MC simulation samples. The default response function
is determined with two photons from $\pizero$ decay and the low energy
($E<100$\,GeV) background particles generated by {\sc epos 1.99}. Validity of
the choice of {\sc epos 1.99} is tested by comparing two corrected spectra, one
generated by {\sc epos 1.99} and another by {\sc pythia 8.145}. No statistically
significant difference between the corrected spectra is found. A chi-square test
of the corrected spectra based on the default response function against the true
spectra ensures the chi-square probability is greater than 60\,\%. Thus it is
concluded that with the background subtraction and unfolding methods used in
this analysis there is no significant bias and the statistical uncertainty is
correctly quoted.
Accordingly no systematic uncertainty related to the choice of the hadronic
interaction models for the reference MC simulations is considered in the
analysis that follows.

% ----- Acceptance -----
\subsection{Acceptance and branching ratio correction}\label{sec:acceptance}

The apertures of the LHCf calorimeters do not cover the full $2\pi$ azimuthal
angle over the entire rapidity range that is sampled. A correction for this
is applied to the data before it is compared with theoretical expectations.

The correction is done using the rapidity -- $\pt$ phase space.
Correction coefficients are determined as follows. First, using a toy MC
simulation, a single $\pizero$ is generated at IP1 and the decay photons are
propagated to the LHCf detectors. The energy-momentum 4-vectors of the
$\pizero$s are randomly chosen so that they cover the rapidity range that the
LHCf detectors are able to measure. The beam pipe shadow on the calorimeter and
the actual detector positions are taken into account using survey data.

Next fiducial area cuts in the transverse X-Y plane are applied to eliminate
particles that do not fall within the acceptance of the calorimeters.
In the fiducial area cuts, a systematic shift of the proton beam axis is applied
according to the reconstruction of the beam-axis during LHC operation.
In addition a cut is applied to eliminate photons with energy less than
100\,GeV. This corresponds to the treatment of the actual data for reducing the
background contamination by particle interactions with the beam pipe.

\begin{figure}[htbp]
  \begin{center}
  \includegraphics[width=8.5cm, keepaspectratio]{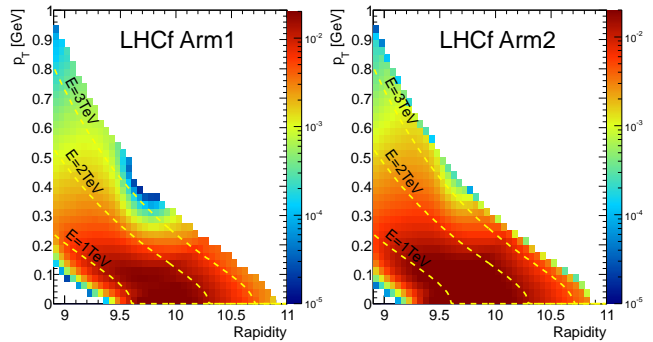}
  \caption{(color online). The acceptance map of $\pizero$ detection by the LHCf
  detectors in rapidity -- $\pt$ phase space: Arm1 (left) and Arm2 (right).
  Fiducial area cuts and energy threshold ($E_{\text{photon}}>100$\,GeV) are
  taken into account. Dashed curves indicate lines of constant energy
  $\pizero$s, $E = 1$\,TeV, 2\,TeV and 3\,TeV.}
  \label{fig:acceptance}
  \end{center}
\end{figure}

Finally two phase space distributions of $\pizero$s are produced; one is for all
$\pizero$s generated at IP1 and the other is for $\pizero$s accepted by the
calorimeters. The ratio of the distribution of accepted $\pizero$s divided by
the distribution of all $\pizero$s is then the geometrical acceptance
efficiency. Fig.~\ref{fig:acceptance} shows the acceptance efficiency as a
function of the $\pizero$ rapidity and $\pt$ and dashed curves indicate lines of
constant $\pizero$ energy, $E = 1$\,TeV, 2\,TeV and 3\,TeV. The left and right
panels indicate the acceptance efficiency for Arm1 and Arm2, respectively.
The final rapidity and $\pt$ spectra are obtained by applying the acceptance map
shown in Fig.~\ref{fig:acceptance} to the acceptance uncorrected data. Note that
the correction maps in Fig.~\ref{fig:acceptance} are purely kinematic and do not
depend upon the particular hadronic interaction model that has been used.
The uncertainty of the acceptance map caused by the finite statistics of the MC
simulations is negligible.

The branching ratio of $\pizero$ decay into two photons is 98.8\,\% and then
inefficiency due to $\pizero$ decay into channels other than two photons
(1.2\,\%) is taken into account by increasing the acceptance efficiency in
rapidity -- $\pt$ phase space by 1.2\,\% everywhere and is independent of the
particular hadronic interaction model.

% ----- Multi-hit pizero -----
\subsection{Multi-hit $\pizero$ correction}\label{sec:multihitpizero}

The detected events have been classified into two types of events: single-hit
$\pizero$ and multi-hit $\pizero$ events.
The former class consists of two photons, one in each of the calorimeters of an
Arm1 or Arm2 detector. A multi-hit $\pizero$ event is defined as a single
$\pizero$ accompanied with at least one additional background particle (photon
or neutron) in one of the calorimeters. In this analysis, only single-hit
$\pizero$ events are considered, and multi-hit $\pizero$ events are rejected in
the single-hit selection process (Sec.~\ref{sec:positionreconstruction}) when
the energy of the additional background particle is beyond the energy threshold
of the cut.

The loss of multi-hit $\pizero$ events is corrected for with the help of event
generators. A range of ratios of multi-hit plus single-hit to single-hit
$\pizero$ events is estimated using several hadronic interaction models in each
rapidity range. The observed $\pt$ spectra are then multiplied by the mean of
these ratios and also contribute a systematic uncertainty corresponding to the
variation among the interaction models. In this way the single-hit $\pizero$
spectra are corrected so they represent inclusive $\pizero$ production spectra.
The $\pt$ dependent range of the flux of multi-hit $\pizero$ events has been
estimated using {\sc qgsjet} II-03, {\sc dpmjet 3.04}, {\sc sibyll 2.1}, {\sc
epos 1.99} and {\sc pythia 8.145}, and resulted in a range of 0\,\%--10\,\% of
the flux of single-hit $\pizero$ events.

% ----- Systematic uncertainty -----
\section{Systematic uncertainties}\label{sec:systerror}

\subsection{Energy scale}\label{sec:syst_energy}

The known rest mass of the $\pizero$s is $134.9766 \pm 0.0006$\,MeV~\cite{PDG}
whereas the peak of the two-photon invariant mass measured by the two LHCf
detectors occurs at 145.8$\pm$0.1\,MeV (Arm1) and 139.9$\pm$0.1\,MeV (Arm2)
where the $\pm$ 0.1\,MeV uncertainties are statistical. The mass excess error is
+8.1\,\% for Arm1 and +3.8\,\% for Arm2. According to Eq.~\ref{eq:pi0mass} there
are two possible sources for mass excess error; (1) systematic over estimates of
the energies $E_1$ and $E_2$ of the two decay photons and (2) systematic over
estimate of the opening angle between the two photons. As discussed in Sec. IV B
the typical uncertainty in opening angle is less than 1\,\%, too small to
explain the observed mass excesses. This leaves measurement of the photon
energies as the source of mass excess error.

The uncertainty in measurement of photon energy has also been investigated in a
beam test at SPS and calibration with a radiation source. The estimated
uncertainty of photon energy from these tests is 3.5\,\%. The 3.5\,\%
uncertainty is dominated by the uncertainties in factors converting measured
charge to deposited energy~\cite{SPS2007}. Note that the linearity of each PMT
was carefully tested before detector assembly over a wide range of signal
amplitude by exciting the scintillator with a 337\,nm UV laser
pulse~\cite{LHCfTDR, prototype}. The difference of reconstructed energy between
the reconstruction algorithm with and without non-linearity correction of PMTs
for 3\,TeV photons is only 0.5\,\% at maximum, nevertheless the measured
non-linear response functions have been applied in the analysis.

The systematic uncertainties estimated by the beam test data at SPS (3.5\,\% for
both Arms) are considered as uncorrelated among the $\pt$ bins, while the
systematic uncertainties owing to the mass excess errors (8.1\,\% for Arm1
and 3.8\,\% for Arm2) are considered as correlated between each $\pt$ bin.
The systematic shift of bin contents due to the energy scale uncertainties is
estimated using two energy spectra by artificially scaling the energy with the
two extremes.
The ratios of the two extreme spectra to the non-scaled spectrum are assigned as
systematic shifts in each bin.

\subsection{Particle identification}\label{sec:syst_pid}

The $L_{90\%}$ distribution described in Sec.~\ref{sec:pid} is used to select
LHCf $\pizero$ events for the $\pt$ spectra presented in Sec.~\ref{sec:result}.
Some disagreements in the $L_{90\%}$ distribution are found between the LHCf
data and the MC simulations. This may be caused by residual errors of the
channel-to-channel calibrations of the LHCf detector relative to the LHCf
detector simulation.

The corresponding systematic uncertainty of the $L_{90\%}$ distribution is
evaluated by comparing the $L_{90\%}$ distribution of the LHCf $\pizero$
candidate events of the measured data with the MC simulation. The $L_{90\%}$
distribution for LHCf $\pizero$ events is increased by at most one radiation
length compared to the MC simulation.
The systematic shifts of $\pt$ spectra bin contents are taken from the ratio of
$\pt$ spectra with artificial shifts of the $L_{90\%}$ distribution to the $\pt$
spectra without any $L_{90\%}$ shift. This effect may distort the measured $\pt$
spectra by 0--20\,\% depending on $\pt$.

\subsection{Offset of beam axis}\label{sec:syst_beamcenter}

In the geometrical analysis of the data, the projected position of the zero
degree collision angle at the LHCf detectors (beam center) can vary from fill to
fill owing to slightly different beam transverse position and crossing angles at
IP1. The beam center at the LHCf detectors can be determined by two methods;
first by using the distribution of particle incident positions measured by the
LHCf detectors and second by using the information from the Beam Position
Monitors (BPMSW) installed $\pm$21\,m from IP1~\cite{BPM}.
Consistent results for the beam center are obtained by the two methods applied
to LHC fills 1089--1134 within 1\,mm accuracy.
The systematic shifts to $\pt$ spectra bin contents are evaluated by taking the
ratio of spectra with the beam-center displaced by 1\,mm to spectra with no
displacement as determined by the distribution of particle incident positions
measured by the LHCf detectors. Owing to the fluctuations of the beam-center
position, the $\pt$ spectra are modified by 5--20\,\% depending on the rapidity
range.

\subsection{Single-hit selection}\label{sec:syst_singlehit}

Since energy reconstruction is degraded when more than one photon hits a given
calorimeter, only single-hit events are used in the analysis. Owing to selection
efficiency greater than 98\,\% for single-hit events and rejection of
contamination by multi-hit events by the invariant mass cut, the systematic
shift caused by the uncertainty in single-hit selection to bin contents is
3\,\%.

\subsection{Position dependent correction}\label{sec:syst_leak}

As described in Sec.~\ref{sec:energyreconstruction}, energy reconstruction of
the photons is sensitive to shower leakage effects which are a function of the
photon incident position. Systematic uncertainties related to the leakage-out
and leakage-in effects arise from residual errors of calorimeter response when
tuning of the LHCf detector simulation to the calibration data taken at
SPS~\cite{SPS2007} that then lead to a mis-reconstruction of energy. Another
source of uncertainties in energy reconstruction is an error in light-yield
collection efficiency which is also dependent on the photon incident position.

The systematic uncertainty due to position dependent effects is estimated by
comparing two distributions of the energy deposited at each incident position
bin. The first distribution is taken from the beam tests at SPS and the second
distribution is generated by toy MC simulations that assume the upstream
geometry of the test beam at SPS. Shifts of reconstructed $\pt$ attributed to
the residual errors in calorimeter response between these two energy
distributions are assigned as the systematic uncertainties. The typical
systematic shifts of Arm1 (Arm2) are 5\,\% (5\,\%) for low $\pt$ and 40\,\%
(30\,\%) for large $\pt$.
Owing to the light guide geometry, the systematic uncertainty of the Arm1
detector is larger than the Arm2 detector.

\subsection{Luminosity}\label{sec:syst_luminosity}

The instantaneous luminosity is derived from the counting rate of the Front
Counters (FC). The calibration of the FC counting rates to the instantaneous
luminosity was made during the Van der Meer scans on April 26th and May 9th
2010~\cite{LHCfLuminosity}.
The calibration factors obtained from two Van der Meer scans differ by 2.1\,\%.
The estimated luminosities by the two FCs for the May 15th data differ by
2.7\,\%. Considering the uncertainty of $\pm$5.0\,\% in the beam intensity
measurement during the Van der Meer scans~\cite{bcnwg-note}, we estimate an
uncertainty of $\pm$6.1\,\% in the luminosity determination.

% ----- Analysis result -----
\section{Results of analysis}\label{sec:result}

\begin{figure*}[htbp]
  \begin{center}
  \includegraphics[width=16cm, keepaspectratio]{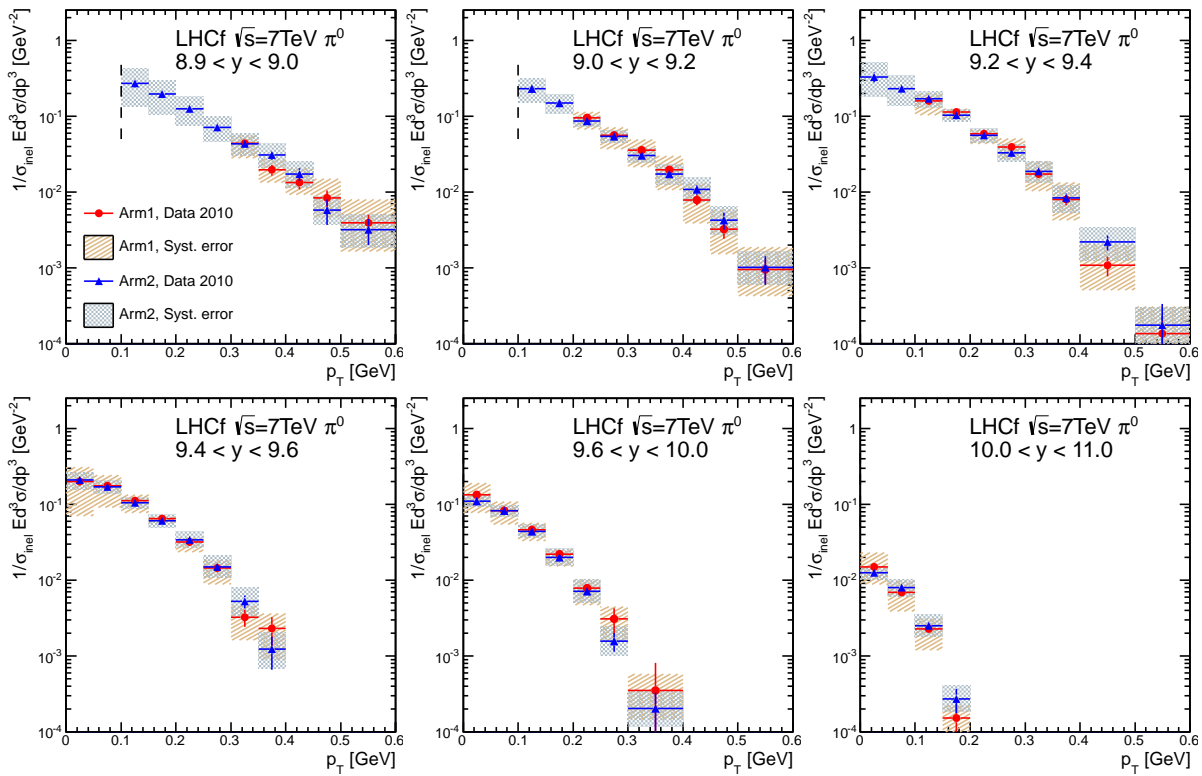}
  \caption{(color online). Experimental $\pt$ spectra of the Arm1 (red dots) and
  Arm2 (blue triangles) detector. Error bars indicate the statistical
  uncertainties and shaded rectangles show the systematic uncertainties of the
  Arm1 and Arm2 detectors.}
  \label{fig:pt_spectra_a1a2}
  \end{center}
\end{figure*}

The $\pt$ spectra derived from the independent analyses of the Arm1 and Arm2
detectors are presented in Fig.~\ref{fig:pt_spectra_a1a2} for six ranges of
rapidity $y$: 8.9 to 9.0, 9.0 to 9.2, 9.2 to 9.4, 9.4 to 9.6, 9.6 to 10.0 and
10.0 to 11.0. The spectra in Fig.~\ref{fig:pt_spectra_a1a2} are after all
corrections discussed in previous sections have been applied. The inclusive
production rate of neutral pions is given by the expression
\begin{equation}
	\frac{1}{\sigma_\text{inel}} E \frac{d^{3}\sigma}{dp^3} =
	\frac{1}{N_\text{inel}}\frac{d^2N(\pt, y)}{2\pi \cdot \pt \cdot d\pt \cdot dy}.
\end{equation}
\noindent $\sigma_\text{inel}$ is the inelastic cross section for proton-proton
collisions at $\sqrt{s} = 7$\,TeV. $E d^{3}\sigma / dp^3$ is the inclusive cross
section of $\pizero$ production. The number of inelastic collisions,
$N_\text{inel}$, used for normalizing the production rates of
Fig.~\ref{fig:pt_spectra_a1a2} has been calculated from $N_\text{inel}$ =
$\sigma_\text{inel} \int {\cal L} dt$, assuming the inelastic cross section
$\sigma_\text{inel}$ = 73.6\,mb. This value for $\sigma_\text{inel}$ has been
derived from the best COMPETE fits~\cite{PDG} and the TOTEM result for the
elastic scattering cross section~\cite{TOTEM}.
Using the integrated luminosities reported in Sec.~\ref{sec:measured_data},
$N_\text{inel}$ is 1.85$\times$10$^{8}$ for Arm1 and 1.40$\times$10$^{8}$ for
Arm2. $d^2N(\pt, y)$ is the number of $\pizero$s detected in the transverse
momentum interval ($d\pt$) and the rapidity interval ($dy$) with all
corrections applied.

In Fig.~\ref{fig:pt_spectra_a1a2}, the red dots and blue triangles represent the
results from Arm1 and Arm2, respectively. The error bars and shaded rectangles
indicate the one standard deviation statistical and total systematic
uncertainties, respectively. The total systematic uncertainties are given by
adding all uncertainty terms except for the luminosity in quadrature.
The vertical dashed lines shown in the rapidity range below 9.2 indicate the
$\pt$ threshold of the Arm2 detector owing to the photon energy threshold and
the geometrical acceptance. The $\pt$ threshold of the Arm1 detector occurs at a
higher value of $\pt$ than Arm2 due to its smaller acceptance.
A general agreement between the Arm1 and Arm2 $\pt$ spectra within statistical
and systematic uncertainties is evident in Fig.~\ref{fig:pt_spectra_a1a2}.

\begin{figure*}[htbp]
  \begin{center}
  \includegraphics[width=16cm, keepaspectratio]{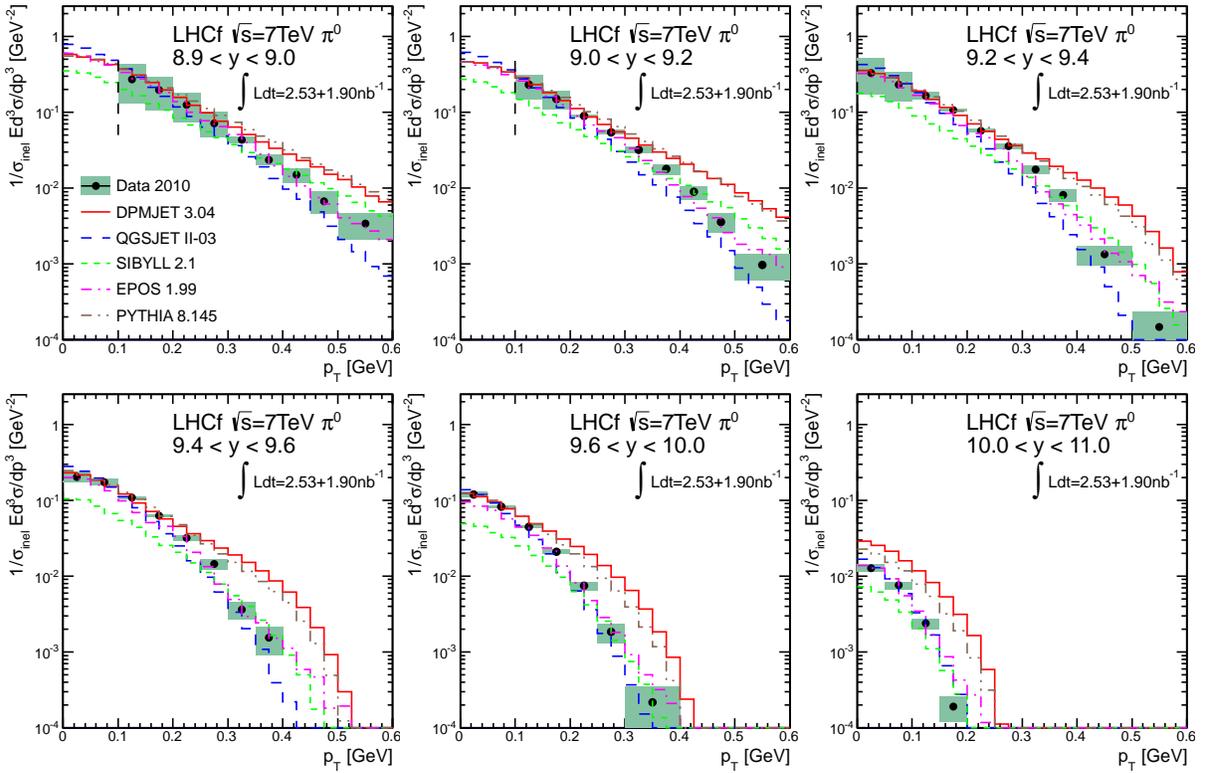}
  \caption{(color online). Combined $\pt$ spectra of the Arm1 and Arm2 detectors
  (black dots) and the total uncertainties (shaded rectangles) compared with the
  predicted spectra by hadronic interaction models.}
  \label{fig:pt_spectra_combined}
  \end{center}
\end{figure*}

Fig.~\ref{fig:pt_spectra_combined} presents the combined $\pt$ spectra of the
Arm1 and Arm2 detectors (black dots). The 68\,\% confidence intervals
incorporating the statistical and systematic uncertainties are indicated by the
shaded green rectangles. The combined spectra below the $\pt$ threshold of Arm1
are taken from the Arm2 spectra alone. Above the $\pt$ threshold of Arm1,
experimental $\pt$ spectra of the Arm1 and Arm2 detectors have been combined
following the ``pull method''~\cite{Pull} and the combined spectra have
accordingly been obtained by minimizing the value of the chi-square function
defined as
  \begin{eqnarray}
	\chi^2 =
	\sum_{i=1}^{n}
	\sum_{a=1}^{2}
	\left(
	\frac{N_{a, i}^\text{obs}
	(1 + \mathit{S}_{a, i}) - N^\text{comb}
	}
	{\sigma_{a, i}}
	\right)^2 
	+ \chi^2_\text{penalty},
	\label{eq:combine_chi2}
  \end{eqnarray}
where the index $i$ represents the $\pt$ bin number running from 1 to $n$ (the
total number of $\pt$ bins), $N_{a, i}^\text{obs}$ is the number of events and
$\sigma_{a, i}$ is the uncertainty of the Arm-$a$ analysis calculated by
quadratically adding the statistical uncertainty and the energy scale
uncertainty estimated by test beam data at SPS. The $\mathit{S}_{a, i}$ denotes
the systematic correction to the number of events in the $i$-th bin of Arm-$a$:
\begin{equation}
	\mathit{S}_{a, i}
	= \sum_{j=1}^{6}f_{a,i}^{j}\varepsilon_{a}^{j}.
	\label{eq:combine_syst}
\end{equation}
The coefficient $f_{a, i}^{j}$ is the systematic shift of $i$-th bin content due
to the $j$-th systematic uncertainty term. The systematic uncertainty is assumed
fully uncorrelated between the Arm1 and Arm2 detectors, and consists of six
uncertainties related to energy scale owing to the invariant mass shift, PID,
beam center position, single-hit, position dependent correction, and
contamination by mulit-hit $\pizero$ events.
Coefficients $\varepsilon_{a}^{j}$, which should follow a Gaussian distribution,
can be varied to achieve the minimum $\chi^2$ value in each chi-square test,
while they are constrained by the penalty term
\begin{equation}
	\chi^2_\text{penalty} =
	\sum_{j=1}^{6}
	\left(
	|\varepsilon_\text{Arm1}^{j}|^2 +
	|\varepsilon_\text{Arm2}^{j}|^2
	\right).
% 	+ |\varepsilon^\text{ene}|^2.
	\label{eq:combine_penalty}
\end{equation}
\noindent The $\pizero$ production rates for the combined data of LHCf are
summarized in Tables~\ref{table:spectra_0}--~\ref{table:spectra_5}.
Note that the uncertainty in the luminosity determination $\pm$6.1\%, that is
not included in Fig.~\ref{fig:pt_spectra_combined}, can make a $\pt$
independent shift of all spectra.

For comparison, the $\pt$ spectra predicted by various hadronic interaction
models are also shown in Fig.~\ref{fig:pt_spectra_combined}. The hadronic
interaction models that have been used in Fig.~\ref{fig:pt_spectra_combined} are
{\sc dpmjet 3.04} (solid, red), {\sc qgsjet} II-03 (dashed, blue), {\sc sibyll
2.1} (dotted, green), {\sc epos 1.99} (dashed dotted, magenta), and {\sc pythia
8.145} (default parameter set, dashed double-dotted, brown). In these MC
simulations, $\pizero$s from short lived particles that decay within 1\,m from
IP1, for example $\eta \to 3\pizero$, are also counted to be consistent with the
treatment of the experimental data. Note that, since the experimental $\pt$
spectra have been corrected for the influences of the detector responses, event
selection efficiencies and geometrical acceptance efficiencies, the $\pt$
spectra of the interaction models may be compared directly to the experimental
spectra as presented in Fig.~\ref{fig:pt_spectra_combined}.

\begin{figure*}[htbp]
  \begin{center}
  \includegraphics[width=16cm, keepaspectratio]{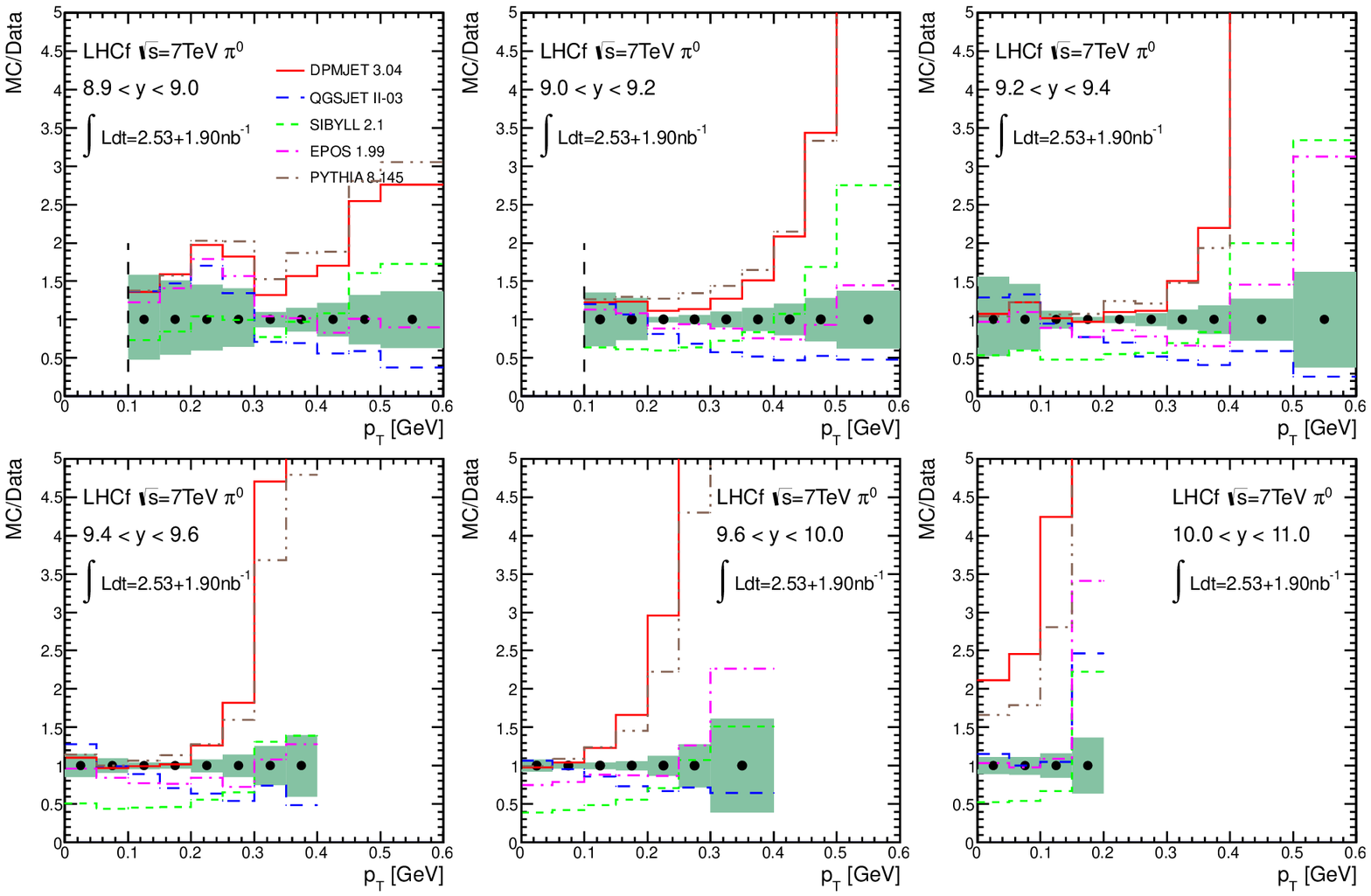}
  \caption{(color online). Ratio of the combined $\pt$ spectra of the Arm1 and
  Arm2 detectors to the predicted $\pt$ spectra by hadronic interaction models.
  Shaded areas indicate the range of total uncertainties of the combined $\pt$
  spectra.}
  \label{fig:pt_spectra_combined_ratio}
  \end{center}
\end{figure*}

Fig.~\ref{fig:pt_spectra_combined_ratio} presents the ratios of $\pt$ spectra
predicted by the various hadronic interaction models to the combined $\pt$
spectra. Error bars have been taken from the statistical and systematic
uncertainties. A slight step found around $\pt = 0.3$\,GeV in $8.9 < y < 9.0$ is
due to low $\pt$ cutoff of the Arm1 data.
The ratios are summarized in
Tables~\ref{table:ratiospectra_0}--~\ref{table:ratiospectra_5}.

% ----- Discussions -----
\section{Discussion}\label{sec:discussions}

\subsection{Transverse momentum spectra}\label{sec:pt_spectra}

Several points can be made about Fig.~\ref{fig:pt_spectra_combined_ratio}.
First, {\sc dpmjet} 3.04 and {\sc pythia} 8.145 show overall agreement with the
LHCf data for $9.2 < y < 9.6$ and $\pt < 0.2$\,GeV, while the expected $\pizero$
production rates by both models exceed the LHCf data as $\pt$ becomes large.
The latter observation can be explained by the baryon/meson production mechanism
that has been employed in both models.
More specifically, the ``popcorn model''~\cite{Andersson, Eden} is used to
produce baryons and mesons through string breaking, and this mechanism tends to
lead to hard pion spectra. {\sc sibyll} 2.1, which is also based on the popcorn
model, also predicts harder pion spectra than the experimental data, although
the expected $\pizero$ yield is generally small.

On the other hand, {\sc qgsjet} II-03 predicts $\pizero$ spectra that are softer
than the LHCf data and the other models. This might be due to the fact that
only one quark exchange is allowed in the {\sc qgsjet} model. The remnants
produced in a proton-proton collision are likewise baryons with relatively small
mass, so fewer pions with large energy are produced.

Among hadronic interaction models tested in this analysis, {\sc epos} 1.99 shows
the best overall agreement with the LHCf data. However {\sc epos} 1.99 behaves
softer than the data in the low $\pt$ region, $\pt \lesssim 0.4$\,GeV in
$9.0 < y <9.4$ and $\pt \lesssim 0.3$\,GeV in $9.4 < y < 9.6$, and behaves
harder in the large $\pt$ region. Specifically a dip found in the ratio of {\sc
epos} 1.99 to the LHCf data for $y > 9.0$ can be attributed to the transition
between two pion production mechanisms: string fragmentation via cut Pomeron
process (low energy $\sim$ low $\pt$ for the fixed rapidity) and remnants of
projectile/target (high energy $\sim$ large $\pt$ for the fixed
rapidity)~\cite{Pierog}.

\subsection{Average transverse momentum}\label{sec:average_pt}

According to the scaling law proposed by several authors~\cite{Amati, Benecke,
Feynman}, the average transverse momentum as a function of rapidity should be
independent of the center of mass energy in the projectile fragmentation region.
Average transverse momentum, $\langle\pt\rangle$, can be obtained by fitting an
empirical function to the $\pt$ spectra in each rapidity range. In this
analysis, among several ansatz proposed for fitting the $\pt$ spectra, an
exponential distribution has been first chosen with the form
\begin{equation}
  \frac{1}{\sigma_\text{inel}}E\frac{d^3\sigma}{dp^3} =
  A\cdot\exp(-\sqrt{{\pt}^2 + m_{\pizero}^2}/T).
  \label{eq:ptfit_exp}
\end{equation}
This distribution is motivated by a thermodynamical model~\cite{Hagedorn}.
The parameter $A$\,[GeV$^{-2}$] is a normalization factor and $T$\,[GeV] is
the temperature of $\pizero$s with a given transverse momentum $\pt$. Using
Eq.~\ref{eq:ptfit_exp}, $\langle\pt\rangle$ is derived as a function of $T$:
\begin{equation}
	\langle\pt\rangle = \sqrt{\frac{\pi m_{\pizero}T}{2}}
	\frac{K_2(m_{\pizero}/T)}{K_{3/2}(m_{\pizero}/T)},
	\label{eq:ptfit_exp_ave}
\end{equation}
where $K_\alpha(m_{\pizero}/T)$ is the modified Bessel function.

\begin{figure}[htbp]
  \begin{center}
  \includegraphics[width=8.8cm, keepaspectratio]{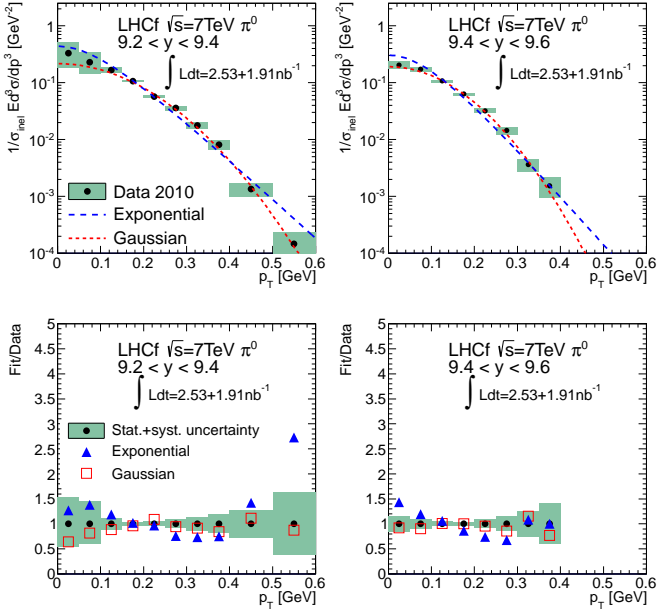}
  \caption{(color online). (Upper) Experimental $\pt$ spectra (black dots and
  green shaded rectangles), the best-fit exponential distributions
  (Eq.~(\ref{eq:ptfit_exp}), dashed curve) and the best-fit Gaussian
  distributions (Eq.~(\ref{eq:ptfit_gauss}), dotted curve).
  (Bottom) Ratios of the best-fit exponential or Gaussian distribution to the
  experimental data (blue triangles or red open boxes) and the statistical and
  systematic uncertainties (green shaded areas). For both the upper and bottom
  panels, the rapidity ranges $9.2 < y < 9.4$ and $9.4 < y < 9.6$ are shown on
  the left and right panels, respectively.}
  \label{fig:pt_combined_fit}
  \end{center}
\end{figure}

Best-fit results for $T$ and $\langle\pt\rangle$ are summarized in
Table~\ref{table:average_pt}.
The worse fit quality values are found for $9.2 < y < 9.4$ ($\chi^2/$dof = 3.6)
and $9.4 < y < 9.6$ ($\chi^2/$dof = 11.1). These are caused by data points near
$\pt = 0.25$\,GeV which exceed the best-fit exponential distribution and the
experimental $\pt$ spectra decreasing more rapidly than
Eq.~(\ref{eq:ptfit_exp}) for $\pt > 0.3$\,GeV. The upper panels in
Fig.~\ref{fig:pt_combined_fit} show the experimental $\pt$ spectra (black dots
and green shaded rectangles) and the best-fit of Eq.~(\ref{eq:ptfit_exp})
(dashed curve) in the rapidity range $9.2 < y < 9.4$ and $9.4 < y < 9.6$. The
bottom panels in Fig.~\ref{fig:pt_combined_fit} show the ratio of the best-fit
distribution to the experimental data (blue triangles). Shaded rectangles
indicate the statistical and systematic uncertainties. Even though the minimum
$\chi^2/$dof values are large, the best-fit $T$ values are consistent
with temperatures that are typical of soft QCD processes and the predictions of
the thermodynamical model ($T \lesssim 180$\,MeV)~\cite{Hagedorn} for $y > 8.9$.

Another possibility is that the $\pt$ distributions in
Fig.~\ref{fig:pt_spectra_combined} can also be described by a Gaussian
distribution:
\begin{equation}
  \frac{1}{\sigma_\text{inel}}E\frac{d^3\sigma}{dp^3} =
  A\frac{\exp(-{\pt}^2/\sigma^2_\text{Gauss})}{\pi\sigma^2_\text{Gauss}}.
  \label{eq:ptfit_gauss}
\end{equation}
\noindent The Gaussian width $\sigma_\text{Gauss}$ determines the mean square
$\pt$ of the $\pt$ spectra. $\langle\pt\rangle$ is derived as a function of
$\sigma_\text{Gauss}$ according to:
\begin{equation}
	\langle\pt\rangle = \cfrac{\int 2\pt^2 f(\pt) d\pt}{\int 2\pt f(\pt) d\pt}
	= \frac{\sqrt{\pi}}{2}\sigma_\text{Gauss}.
	\label{eq:ptfit_gauss_ave}
\end{equation}
\noindent where $f(\pt)$ is given by Eq.~(\ref{eq:ptfit_gauss}). Best-fit
results for $\sigma_\text{Gauss}$ and $\langle\pt\rangle$ are summarized in
Table~\ref{table:average_pt}. In this case good fit quality values are found for
all rapidity ranges. The best-fit of Eq.~(\ref{eq:ptfit_gauss}) (dotted curve)
and the ratio of the best-fit Gaussian distribution to the experimental data
(red open boxes) are found in Fig.~\ref{fig:pt_combined_fit}.

A third approach for estimating $\langle\pt\rangle$ is simply numerically
integrating the $\pt$ spectra. With this approach $\langle\pt\rangle$ is given
by
\begin{equation}
	\langle\pt\rangle =
	\cfrac{
	\int_0^{\infty}2\pi\pt^2 f(\pt) d\pt}{
	\int_0^{\infty}2\pi\pt   f(\pt) d\pt}.
\end{equation}
\noindent where $f(\pt)$ is the measured spectrum given in
Fig.~\ref{fig:pt_spectra_combined} for each of the six ranges of rapidity. In
this analysis, $\langle\pt\rangle$ is obtained over the rapidity range
$9.2<y<11.0$ where the $\pt$ spectra are available down to 0\,GeV.
Although the upper limits of numerical integration are actually finite,
$\pt^\text{upper} \leqq 0.6$\,GeV, the contribution of the high $\pt$ tail to
$\langle\pt\rangle$ is negligible. $\pt^\text{upper}$ and the obtained
$\langle\pt\rangle$ are summarized in Table~\ref{table:average_pt}.

The values of $\langle\pt\rangle$ obtained by the three methods discussed above
are in general agreement. When a specific values of $\langle\pt\rangle$ are
needed for this paper the values chosen ($\langle\pt\rangle_\text{LHCf}$) are
defined as follows.
For the rapidity range $8.9<y<9.2$, $\langle\pt\rangle_\text{LHCf}$ is taken
from the weighted mean of $\langle\pt\rangle$ obtained by the exponential fit of
Eq.~(\ref{eq:ptfit_exp_ave}) and the Gaussian fit of
Eq.~(\ref{eq:ptfit_gauss_ave}). The systematic uncertainty related to a possible
bias of the $\langle\pt\rangle$ extraction methods is estimated by the
difference of $\langle\pt\rangle$ derived from these two different fitting
functions. The estimated systematic uncertainty is $\pm6$\,\% for both rapidity
bins.
For the rapidity range $9.2<y<11.0$, the results obtained by the Gaussian fit
and numerical integration are used to calculate the weighted mean of
$\langle\pt\rangle_\text{LHCf}$ in order to avoid the poor quality of fit of the
exponential function in this rapidity range. Systematic uncertainty is estimated
to be $\pm$3\,\% and $\pm$2\,\% for $9.2<y<9.4$ and $9.4<y<11.0$, respectively.
The values of $\langle\pt\rangle_\text{LHCf}$ obtained by the above calculation
are summarized in Table~\ref{table:average_pt_final}.

\begin{widetext}
\begin{table*}[htbp]
	\begin{center}
	\begin{tabular}{l || r c c c | r c c c | c c c} \hline
    \multicolumn{1}{l ||}{~} &
    \multicolumn{4}{c|}{Exponential fit} &
    \multicolumn{4}{c|}{Gaussian fit}    &
    \multicolumn{3}{c }{Numerical integration}   \\
    \hline
    Rapidity         & $\chi^2$ (dof)
    & $T$ & $\langle\pt\rangle$ & Stat. error & $\chi^2$ (dof) & $\sigma_\text{Gauss}$ & $\langle\pt\rangle$& Stat. error
                                      & $\pt^\text{upper}$ &$\langle\pt\rangle$& Stat. error\\
	                 &                & [MeV] & [MeV] & [MeV]
	                 &                & [MeV] & [MeV] & [MeV]
	                                  & [GeV] & [MeV] & [MeV] \\
	\hline
	$[$8.9,  9.0$]$  &   0.6 (7)  & 83.8  & 201.4 & 13.5
	 				 &   2.0 (7)  & 259.0 & 229.6 & 13.1 
	 				              &       &       &      \\
	$[$9.0,  9.2$]$  &   8.2 (7)  & 75.2  & 184.1 & 5.0 
					 &   0.9 (7)  & 234.7 & 208.0 & 4.6 
					              &       &       &      \\
	$[$9.2,  9.4$]$  &  28.7 (8)  & 61.7  & 164.0 & 2.8  
	                 &   6.9 (8)  & 201.8 & 178.9 & 3.4
	                              &   0.6 & 167.7 & 9.6  \\
	$[$9.4,  9.6$]$  &  66.3 (6)  & 52.8  & 140.3 & 1.9
	                 &   3.3 (6)  & 166.3 & 147.4 & 2.7 
	                              &   0.4 & 144.8 & 3.2  \\
	$[$9.6,  10.0$]$ &  14.0 (5)  & 43.3  & 123.5 & 2.2
	                 &   0.3 (5)  & 139.2 & 123.3 & 3.0 
	                              &   0.4 & 117.0 & 2.1  \\
	$[$10.0, 11.0$]$ &   9.0 (2)  & 21.3  &  77.7 & 2.3
	                 &   2.1 (2)  &  84.8 &  75.1 & 2.9
	                              &   0.2 &  76.9 & 2.6  \\
	\hline
	
	\end{tabular}
	\caption{Best-fit results of exponential and Gaussian $\pt$ functions to the
	LHCf data and average $\pizero$ transverse momenta for the rapidity range
	8.9$<$y$<$11.0 obtained by using the exponential fit, Gaussian fit and
	numerical integration.
	\label{table:average_pt}}
	\end{center}
\end{table*}
\end{widetext}

% \begin{table}[htbp]
% 	\begin{center}
% 	\begin{tabular}{l r c c c}
% 	\hline
% Rapidity & $\chi^2$ (dof) & $\sigma_\text{Gauss}$ & $\langle\pt\rangle$ & Stat.
% uncertainty
% \\
% 	     &                & [MeV] & [MeV]         & [MeV]         \\
% 	\hline
% 	$[$8.9,  9.0$]$  &   2.0 (7)  & 259.0 & 229.6 & 13.1 \\
% 	$[$9.0,  9.2$]$  &   0.9 (7)  & 234.7 & 208.0 & 4.6  \\
% 	$[$9.2,  9.4$]$  &   6.9 (8)  & 201.8 & 178.9 & 3.4  \\
% 	$[$9.4,  9.6$]$  &   3.3 (6)  & 166.3 & 147.4 & 2.7  \\
% 	$[$9.6,  10.0$]$ &   0.3 (5)  & 139.2 & 123.3 & 3.0  \\
% 	$[$10.0, 11.0$]$ &   2.1 (2)  &  84.8 &  75.1 & 2.9  \\
% 	\hline
% 	
% 	\end{tabular}
% \caption{Best-fit results of the fitting a Gaussian function to the LHCf
% data and average transverse momentum of $\pizero$ for the rapidity range
% 8.9$<$y$<$11.0.
% 	\label{table:average_pt3}}
% 	\end{center}
% \end{table}
% 
% \begin{table}[htbp]
% 	\begin{center}
% 	\begin{tabular}{l c c c}
% 	\hline
% 	Rapidity & $\pt^\text{upper}$ & $\langle\pt\rangle$ & Stat. uncertainty \\
% 	         & [GeV]          & [MeV]           & [MeV]        \\
% 	\hline
% 	$[$9.2, 9.4$]$   & 0.6 & 167.7 & 9.6 \\
% 	$[$9.4, 9.6$]$   & 0.4 & 144.8 & 3.2 \\
% 	$[$9.6, 10.0$]$  & 0.4 & 117.0 & 2.1 \\
% 	$[$10.0, 11.0$]$ & 0.2 &  76.9 & 2.6 \\
% 	\hline
% 	
% 	\end{tabular}
% \caption{Average transverse momentum of $\pizero$ derived by numerical
% integration of the $\pt$ spectra for the rapidity range 9.2$<$y$<$11.0.
% 	\label{table:average_pt2}}
% 	\end{center}
% \end{table}

\begin{table}[htbp]
	\begin{center}
	\begin{tabular}{l c c}
	\hline
Rapidity &             $\langle\pt\rangle$ & Total uncertainty \\
	     &               [MeV]             &     [MeV]         \\
	\hline
	$[$8.9,  9.0$]$  &  215.3 & 17.3 \\
	$[$9.0,  9.2$]$  &  196.8 & 12.5 \\
	$[$9.2,  9.4$]$  &  172.2 & 5.9  \\
	$[$9.4,  9.6$]$  &  146.3 & 3.9  \\
	$[$9.6,  10.0$]$ &  119.2 & 3.4  \\
	$[$10.0, 11.0$]$ &   75.8 & 2.9  \\
	\hline
	
	\end{tabular}
	\caption{Average transverse momentum of $\pizero$ for the rapidity range
	8.9$<$y$<$11.0. Total $\pt$ uncertainty includes both the statistical and
	systematic uncertainties.
	\label{table:average_pt_final}}
	\end{center}
\end{table}

\begin{figure}[htbp]
  \begin{center}
  \includegraphics[width=7cm, keepaspectratio]{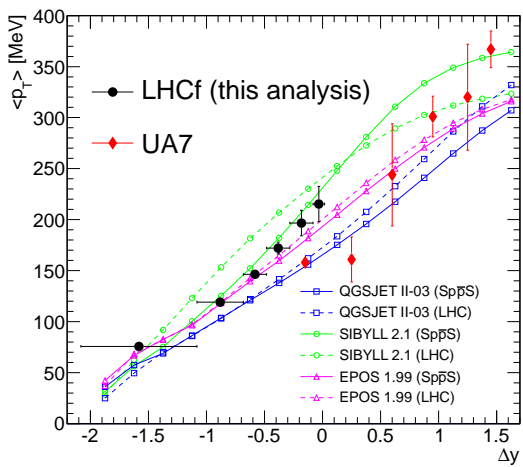}
  \caption{(color online). Average $\pt$ as a function of rapidity loss $\Delta
  y$. Black dots and red diamonds indicate the LHCf data and UA7 results taken
  from Ref.~\cite{UA7}, respectively. The predictions of hadronic interaction
  models are shown by open boxes ({\sc sibyll} 2.1), open circles ({\sc qgsjet}
  II-03) and open triangles ({\sc epos} 1.99). For the predictions of the three
  models, solid and dashed curves indicate the results for the center of mass
  energy at the Sp$\bar{\text{p}}$S and the LHC, respectively.}
  \label{fig:average_pt}
  \end{center}
\end{figure}

The values of $\langle\pt\rangle$ that have been obtained in this analysis,
shown in Table~\ref{table:average_pt_final}, are compared in
Fig.~\ref{fig:average_pt} with the results from UA7 at Sp$\bar{\text{p}}$S
($\sqrt{s} = 630$\,GeV)~\cite{UA7} and the predictions of several hadronic
interaction models. In Fig.~\ref{fig:average_pt} $\langle\pt\rangle$ is
presented as a function of rapidity loss $\Delta y \equiv y_\text{beam} - y$,
where beam rapidity $y_\text{beam}$ is 8.92 for $\sqrt{s} = 7$\,TeV and 6.50 for
$\sqrt{s} = 630$\,GeV. This shift of rapidity scales the results with beam
energy and it allows a direct comparison between LHCf results and past
experimental results at different collision energies. The black dots and the red
diamonds indicate the LHCf data and the UA7 results, respectively.
Although the LHCf and UA7 data in Fig.~\ref{fig:average_pt} have limited overlap
and the systematic errors of the UA7 data are relatively large, the
$\langle\pt\rangle$ spectra for LHCf and UA7 in Fig.~\ref{fig:average_pt} mostly
appear to lie along a common curve.

The $\langle\pt\rangle$ predicted by hadronic interaction models are shown by
open circles ({\sc sibyll} 2.1), open boxes ({\sc qgsjet} II-03) and open
triangles ({\sc epos} 1.99). {\sc sibyll} 2.1 typically gives harder $\pizero$
spectra (larger $\langle\pt\rangle$) and {\sc qgsjet} II-03 gives softer
$\pizero$ spectra (smaller $\langle\pt\rangle$) than the experimental data. For
each prediction, solid and dashed lines indicate $\langle\pt\rangle$ at the
center of mass energy at Sp$\bar{\text{p}}$S and the LHC, respectively.
Of the three models the predictions by {\sc epos} 1.99 show the smallest
dependence of $\langle\pt\rangle$ on the two center of mass energies, and this
tendency is consistent with the LHCf and UA7 results except for the UA7 data at
$\Delta y = -0.15$ and 0.25. It is also evident in Fig.~\ref{fig:average_pt}
that amongst the three models the best agreement with the LHCf data is obtained
by {\sc epos} 1.99.

% ----- Summary -----
\section{Conclusions}\label{sec:conclusions}

The inclusive production of neutral pions in the rapidity range larger than $y =
8.9$ has been measured by the LHCf experiment in proton-proton collisions at the
LHC in early 2010. Transverse momentum spectra of neutral pions have been
measured by two independent LHCf detectors, Arm1 and Arm2, and give consistent
results.
The combined Arm1 and Arm2 spectra have been compared with the predictions of
several hadronic interaction models. {\sc dpmjet} 3.04, {\sc epos} 1.99 and {\sc
pythia} 8.145 agree with the LHCf combined results in general for the rapidity
range $9.0 < y < 9.6$ and $\pt < 0.2$\,GeV. {\sc qgsjet} II-03 has poor
agreement with LHCf data for $8.9 < y < 9.4$, while it agrees with LHCf data for
$y > 9.4$. Among the hadronic interaction models tested in this paper, {\sc
epos} 1.99 shows the best overall agreement with the LHCf data even for $y >
9.6$.

The average transverse momentum, $\langle\pt\rangle$, of the combined $\pt$
spectra is consistent with typical values for soft QCD processes.
The $\langle\pt\rangle$ spectra for LHCf and UA7 in Fig.~\ref{fig:average_pt}
mostly appear to lie along a common curve. The $\langle\pt\rangle$ spectra
derived by LHCf agrees with the expectation of {\sc epos} 1.99.
Additional experimental data are needed to establish the dependence, or
independence, of $\langle\pt\rangle$ on the center of mass collision energy.

% ----- Acknowledgements -----
\section*{Acknowledgments}\label{sec:acknowledgements}

We thank the CERN staff and the ATLAS collaboration for their essential
contributions to the successful operation of LHCf. We also thank Tanguy Pierog
for numerous discussions. This work is partly supported by Grant-in-Aid for
Scientific research by MEXT of Japan and by the Grant-in-Aid for Nagoya
University GCOE "QFPU" from MEXT. This work is also supported by Istituto
Nazionale di Fisica Nucleare (INFN) in Italy.
A part of this work was performed using the computer resource provided by the
Institute for the Cosmic-Ray Research (ICRR), University of Tokyo.

% ----- Appendix -----
\section*{Appendix}\label{sec:appendix}

% Relationship between $\pizero$ spectra and photon epectra is discussed.
% Fig.~\ref{fig:parentpi0_loweta} and Fig.~\ref{fig:parentpi0_highweta} indicate
% the fraction of $\pizero$ events that contribute to the photon energy spectra at
% the $\sqrt{s}=7$ TeV proton-proton collisions for the lower and higher rapidity
% range, respectively.
% 
% \begin{figure*}[htbp]
%   \begin{center}
%   \includegraphics[width=16cm,
%   keepaspectratio]{parentPi0_LowEta_20120206_v2.eps}
%   \caption{(color online).}
%   \label{fig:parentpi0_loweta}
%   \end{center}
% \end{figure*}
% 
% \begin{figure*}[htbp]
%   \begin{center}
%   \includegraphics[width=16cm,
%   keepaspectratio]{parentPi0_HighEta_20120206_v2.eps}
%   \caption{(color online).}
%   \label{fig:parentpi0_highweta}
%   \end{center}
% \end{figure*}

%%-------------------------------------
%% Inclusive pizero production rate
%%-------------------------------------
The inclusive production rates of $\pizero$s measured by LHCf are summarized in
Tables~\ref{table:spectra_0}--~\ref{table:spectra_5}.
The ratios of inclusive production rates of $\pizero$s predicted by MC
simulations to the LHCf measurements are summarized in
Tables~\ref{table:ratiospectra_0}--~\ref{table:ratiospectra_5}.

\begin{table*}[htbp]
	\begin{center}
	\begin{tabular*}{15cm}{@{\extracolsep{\fill}}|c c c c|}
	\hline
	$\pt$ range & Production rate
	& Stat. uncertainty
	& Syst.+Stat. uncertainty \\
	$[$GeV$]$ & $[$GeV$^{-2}]$
	& $[$GeV$^{-2}]$
	& $[$GeV$^{-2}]$ \\
	\hline
$[$0.10, 0.15$]$ & 2.71$\times10^{-1}$ & $\pm$1.41$\times10^{-1}$ & -1.41$\times10^{-1}$, +1.58$\times10^{-1}$ \\
$[$0.15, 0.20$]$ & 1.95$\times10^{-1}$ & $\pm$8.85$\times10^{-2}$ & -8.85$\times10^{-2}$, +9.95$\times10^{-2}$ \\
$[$0.20, 0.25$]$ & 1.25$\times10^{-1}$ & $\pm$4.98$\times10^{-2}$ & -4.98$\times10^{-2}$, +5.66$\times10^{-2}$ \\
$[$0.25, 0.30$]$ & 7.15$\times10^{-2}$ & $\pm$2.54$\times10^{-2}$ & -2.54$\times10^{-2}$, +2.90$\times10^{-2}$ \\
$[$0.30, 0.35$]$ & 4.34$\times10^{-2}$ & $\pm$3.21$\times10^{-3}$ & -4.21$\times10^{-3}$, +4.22$\times10^{-3}$ \\
$[$0.35, 0.40$]$ & 2.36$\times10^{-2}$ & $\pm$2.45$\times10^{-3}$ & -3.65$\times10^{-3}$, +3.66$\times10^{-3}$ \\
$[$0.40, 0.45$]$ & 1.50$\times10^{-2}$ & $\pm$2.05$\times10^{-3}$ & -3.25$\times10^{-3}$, +3.26$\times10^{-3}$ \\
$[$0.45, 0.50$]$ & 6.73$\times10^{-3}$ & $\pm$1.50$\times10^{-3}$ & -2.15$\times10^{-3}$, +2.16$\times10^{-3}$ \\
$[$0.50, 0.60$]$ & 3.42$\times10^{-3}$ & $\pm$8.08$\times10^{-4}$ & -1.27$\times10^{-3}$, +1.27$\times10^{-3}$ \\
	\hline
	\end{tabular*}
	\caption{Production rate for the $\pizero$ production in the rapidity
	range 8.9$<$y$<$9.0. \label{table:spectra_0}}
	\end{center}
\end{table*}

\begin{table*}[htbp]
	\begin{center}
	\begin{tabular*}{15cm}{@{\extracolsep{\fill}}|c c c c|}
	\hline
	$\pt$ range & Production rate
	& Stat. uncertainty
	& Syst.+Stat. uncertainty \\
	$[$GeV$]$ & $[$GeV$^{-2}]$
	& $[$GeV$^{-2}]$
	& $[$GeV$^{-2}]$ \\
	\hline
$[$0.10, 0.15$]$ & 2.30$\times10^{-1}$ & $\pm$8.06$\times10^{-2}$ & -8.06$\times10^{-2}$, +8.12$\times10^{-2}$ \\
$[$0.15, 0.20$]$ & 1.51$\times10^{-1}$ & $\pm$3.99$\times10^{-2}$ & -3.99$\times10^{-2}$, +4.19$\times10^{-2}$ \\
$[$0.20, 0.25$]$ & 8.92$\times10^{-2}$ & $\pm$2.62$\times10^{-3}$ & -3.14$\times10^{-3}$, +3.14$\times10^{-3}$ \\
$[$0.25, 0.30$]$ & 5.43$\times10^{-2}$ & $\pm$2.18$\times10^{-3}$ & -3.35$\times10^{-3}$, +3.35$\times10^{-3}$ \\
$[$0.30, 0.35$]$ & 3.21$\times10^{-2}$ & $\pm$1.79$\times10^{-3}$ & -3.31$\times10^{-3}$, +3.32$\times10^{-3}$ \\
$[$0.35, 0.40$]$ & 1.80$\times10^{-2}$ & $\pm$1.44$\times10^{-3}$ & -2.72$\times10^{-3}$, +2.73$\times10^{-3}$ \\
$[$0.40, 0.45$]$ & 8.91$\times10^{-3}$ & $\pm$1.00$\times10^{-3}$ & -1.83$\times10^{-3}$, +1.84$\times10^{-3}$ \\
$[$0.45, 0.50$]$ & 3.59$\times10^{-3}$ & $\pm$6.49$\times10^{-4}$ & -1.01$\times10^{-3}$, +1.01$\times10^{-3}$ \\
$[$0.50, 0.60$]$ & 9.72$\times10^{-4}$ & $\pm$2.65$\times10^{-4}$ & -3.66$\times10^{-4}$, +3.65$\times10^{-4}$ \\
	\hline
	\end{tabular*}
	\caption{Production rate for the $\pizero$ production in the rapidity
	range 9.0$<$y$<$9.2. \label{table:spectra_1}}
	\end{center}
\end{table*}

\begin{table*}[htbp]
	\begin{center}
	\begin{tabular*}{15cm}{@{\extracolsep{\fill}}|c c c c|}
	\hline
	$\pt$ range & Production rate
	& Stat. uncertainty
	& Syst.+Stat. uncertainty \\
	$[$GeV$]$ & $[$GeV$^{-2}]$
	& $[$GeV$^{-2}]$
	& $[$GeV$^{-2}]$ \\
	\hline
$[$0.00, 0.05$]$ & 3.31$\times10^{-1}$ & $\pm$1.58$\times10^{-1}$ & -1.58$\times10^{-1}$, +1.84$\times10^{-1}$ \\
$[$0.05, 0.10$]$ & 2.31$\times10^{-1}$ & $\pm$9.05$\times10^{-2}$ & -9.05$\times10^{-2}$, +1.07$\times10^{-1}$ \\
$[$0.10, 0.15$]$ & 1.66$\times10^{-1}$ & $\pm$3.23$\times10^{-3}$ & -1.96$\times10^{-2}$, +1.96$\times10^{-2}$ \\
$[$0.15, 0.20$]$ & 1.06$\times10^{-1}$ & $\pm$2.42$\times10^{-3}$ & -3.76$\times10^{-3}$, +3.76$\times10^{-3}$ \\
$[$0.20, 0.25$]$ & 5.71$\times10^{-2}$ & $\pm$1.91$\times10^{-3}$ & -2.69$\times10^{-3}$, +2.69$\times10^{-3}$ \\
$[$0.25, 0.30$]$ & 3.58$\times10^{-2}$ & $\pm$1.65$\times10^{-3}$ & -2.97$\times10^{-3}$, +2.98$\times10^{-3}$ \\
$[$0.30, 0.35$]$ & 1.77$\times10^{-2}$ & $\pm$1.26$\times10^{-3}$ & -2.29$\times10^{-3}$, +2.29$\times10^{-3}$ \\
$[$0.35, 0.40$]$ & 8.07$\times10^{-3}$ & $\pm$9.02$\times10^{-4}$ & -1.49$\times10^{-3}$, +1.49$\times10^{-3}$ \\
$[$0.40, 0.50$]$ & 1.35$\times10^{-3}$ & $\pm$2.66$\times10^{-4}$ & -3.71$\times10^{-4}$, +3.72$\times10^{-4}$ \\
$[$0.50, 0.60$]$ & 1.47$\times10^{-4}$ & $\pm$8.16$\times10^{-5}$ & -9.17$\times10^{-5}$, +9.18$\times10^{-5}$ \\
	\hline
	\end{tabular*}
	\caption{Production rate for the $\pizero$ production in the rapidity
	range 9.2$<$y$<$9.4. \label{table:spectra_2}}
	\end{center}
\end{table*}

\begin{table*}[htbp]
	\begin{center}
	\begin{tabular*}{15cm}{@{\extracolsep{\fill}}|c c c c|}
	\hline
	$\pt$ range & Production rate
	& Stat. uncertainty
	& Syst.+Stat. uncertainty \\
	$[$GeV$]$ & $[$GeV$^{-2}]$
	& $[$GeV$^{-2}]$
	& $[$GeV$^{-2}]$ \\
	\hline
$[$0.00, 0.05$]$ & 2.03$\times10^{-1}$ & $\pm$8.63$\times10^{-3}$ & -3.09$\times10^{-2}$, +3.09$\times10^{-2}$ \\
$[$0.05, 0.10$]$ & 1.73$\times10^{-1}$ & $\pm$3.75$\times10^{-3}$ & -1.64$\times10^{-2}$, +1.64$\times10^{-2}$ \\
$[$0.10, 0.15$]$ & 1.07$\times10^{-1}$ & $\pm$2.24$\times10^{-3}$ & -4.61$\times10^{-3}$, +4.60$\times10^{-3}$ \\
$[$0.15, 0.20$]$ & 6.30$\times10^{-2}$ & $\pm$1.80$\times10^{-3}$ & -2.45$\times10^{-3}$, +2.45$\times10^{-3}$ \\
$[$0.20, 0.25$]$ & 3.20$\times10^{-2}$ & $\pm$1.51$\times10^{-3}$ & -2.63$\times10^{-3}$, +2.64$\times10^{-3}$ \\
$[$0.25, 0.30$]$ & 1.45$\times10^{-2}$ & $\pm$1.17$\times10^{-3}$ & -2.14$\times10^{-3}$, +2.15$\times10^{-3}$ \\
$[$0.30, 0.35$]$ & 3.64$\times10^{-3}$ & $\pm$6.44$\times10^{-4}$ & -9.28$\times10^{-4}$, +9.29$\times10^{-4}$ \\
$[$0.35, 0.40$]$ & 1.54$\times10^{-3}$ & $\pm$4.88$\times10^{-4}$ & -6.20$\times10^{-4}$, +6.21$\times10^{-4}$ \\
$[$0.40, 0.50$]$ & 5.43$\times10^{-5}$ & $\pm$6.19$\times10^{-5}$ & -6.41$\times10^{-5}$, +6.41$\times10^{-5}$ \\
	\hline
	\end{tabular*}
	\caption{Production rate for the $\pizero$ production in the rapidity
	range 9.4$<$y$<$9.6. \label{table:spectra_3}}
	\end{center}
\end{table*}

\begin{table*}[htbp]
	\begin{center}
	\begin{tabular*}{15cm}{@{\extracolsep{\fill}}|c c c c|}
	\hline
	$\pt$ range & Production rate
	& Stat. uncertainty
	& Syst.+Stat. uncertainty \\
	$[$GeV$]$ & $[$GeV$^{-2}]$
	& $[$GeV$^{-2}]$
	& $[$GeV$^{-2}]$ \\
	\hline
$[$0.00, 0.05$]$ & 1.20$\times10^{-1}$ & $\pm$3.49$\times10^{-3}$ & -9.66$\times10^{-3}$, +9.68$\times10^{-3}$ \\
$[$0.05, 0.10$]$ & 8.28$\times10^{-2}$ & $\pm$1.55$\times10^{-3}$ & -2.89$\times10^{-3}$, +2.90$\times10^{-3}$ \\
$[$0.10, 0.15$]$ & 4.49$\times10^{-2}$ & $\pm$1.02$\times10^{-3}$ & -1.88$\times10^{-3}$, +1.88$\times10^{-3}$ \\
$[$0.15, 0.20$]$ & 2.10$\times10^{-2}$ & $\pm$8.40$\times10^{-4}$ & -1.28$\times10^{-3}$, +1.28$\times10^{-3}$ \\
$[$0.20, 0.25$]$ & 7.43$\times10^{-3}$ & $\pm$6.05$\times10^{-4}$ & -9.73$\times10^{-4}$, +9.76$\times10^{-4}$ \\
$[$0.25, 0.30$]$ & 1.84$\times10^{-3}$ & $\pm$4.05$\times10^{-4}$ & -5.15$\times10^{-4}$, +5.16$\times10^{-4}$ \\
$[$0.30, 0.40$]$ & 2.17$\times10^{-4}$ & $\pm$1.21$\times10^{-4}$ & -1.33$\times10^{-4}$, +1.33$\times10^{-4}$ \\
	\hline
	\end{tabular*}
	\caption{Production rate for the $\pizero$ production in the rapidity
	range 9.6$<$y$<$10.0. \label{table:spectra_4}}
	\end{center}
\end{table*}

\begin{table*}[htbp]
	\begin{center}
	\begin{tabular*}{15cm}{@{\extracolsep{\fill}}|c c c c|}
	\hline
	$\pt$ range & Production rate
	& Stat. uncertainty
	& Syst.+Stat. uncertainty \\
	$[$GeV$]$ & $[$GeV$^{-2}]$
	& $[$GeV$^{-2}]$
	& $[$GeV$^{-2}]$ \\
	\hline
$[$0.00, 0.05$]$ & 1.28$\times10^{-2}$ & $\pm$9.69$\times10^{-4}$ & -1.42$\times10^{-3}$, +1.42$\times10^{-3}$ \\
$[$0.05, 0.10$]$ & 7.55$\times10^{-3}$ & $\pm$3.79$\times10^{-4}$ & -8.88$\times10^{-4}$, +8.85$\times10^{-4}$ \\
$[$0.10, 0.15$]$ & 2.37$\times10^{-3}$ & $\pm$1.95$\times10^{-4}$ & -3.77$\times10^{-4}$, +3.76$\times10^{-4}$ \\
$[$0.15, 0.20$]$ & 1.91$\times10^{-4}$ & $\pm$6.22$\times10^{-5}$ & -6.99$\times10^{-5}$, +6.98$\times10^{-5}$ \\
$[$0.20, 0.30$]$ & 8.37$\times10^{-6}$ & $\pm$1.03$\times10^{-5}$ & -1.05$\times10^{-5}$, +1.05$\times10^{-5}$ \\
	\hline
	\end{tabular*}
	\caption{Production rate for the $\pizero$ production in the rapidity
	range 10.0$<$y$<$11.0. \label{table:spectra_5}}
	\end{center}
\end{table*}

%%-------------------------------------
%% Ratio of the MC simulations to data
%%-------------------------------------

\begin{table*}[htbp]
	\begin{center}
	\begin{tabular*}{12cm}{@{\extracolsep{\fill}}|c c c c c c|}
	\hline
	$\pt$ range & {\sc dpmjet} 3.04 & {\sc qgsjet} II-03 & {\sc sibyll} 2.1 & {\sc
	epos} 1.99 & {\sc pythia} 8.145 \\
	$[$GeV$]$ & & & & & \\
	\hline
$[$0.10, 0.15$]$ & 1.36 & 1.37 & 0.74 & 1.23 & 1.38 \\
$[$0.15, 0.20$]$ & 1.59 & 1.48 & 0.85 & 1.41 & 1.57 \\
$[$0.20, 0.25$]$ & 1.97 & 1.71 & 1.04 & 1.79 & 2.03 \\
$[$0.25, 0.30$]$ & 1.82 & 1.34 & 1.00 & 1.57 & 2.02 \\
$[$0.30, 0.35$]$ & 1.32 & 0.71 & 0.77 & 1.04 & 1.53 \\
$[$0.35, 0.40$]$ & 1.57 & 0.69 & 0.97 & 1.02 & 1.87 \\
$[$0.40, 0.45$]$ & 1.70 & 0.56 & 1.08 & 0.83 & 1.89 \\
$[$0.45, 0.50$]$ & 2.54 & 0.59 & 1.60 & 1.01 & 2.81 \\
$[$0.50, 0.60$]$ & 2.76 & 0.38 & 1.73 & 0.90 & 3.05 \\
	\hline
	\end{tabular*}
	\caption{Ratio of $\pizero$ production rate of MC simulation to data in the
	rapidity range 8.9$<$y$<$9.0. \label{table:ratiospectra_0}}
	\end{center}
\end{table*}
	
\begin{table*}[htbp]
	\begin{center}
	\begin{tabular*}{12cm}{@{\extracolsep{\fill}}|c c c c c c|}
	\hline
	$\pt$ range & {\sc dpmjet} 3.04 & {\sc qgsjet} II-03 & {\sc sibyll} 2.1 & {\sc
	epos} 1.99 & {\sc pythia} 8.145 \\
	$[$GeV$]$ & & & & & \\
	\hline
$[$0.10, 0.15$]$ & 1.23 & 1.20 & 0.64 & 1.13 & 1.26 \\
$[$0.15, 0.20$]$ & 1.23 & 1.06 & 0.62 & 1.09 & 1.30 \\
$[$0.20, 0.25$]$ & 1.11 & 0.81 & 0.60 & 0.88 & 1.28 \\
$[$0.25, 0.30$]$ & 1.14 & 0.68 & 0.64 & 0.94 & 1.34 \\
$[$0.30, 0.35$]$ & 1.27 & 0.58 & 0.73 & 0.88 & 1.44 \\
$[$0.35, 0.40$]$ & 1.51 & 0.52 & 0.84 & 0.76 & 1.65 \\
$[$0.40, 0.45$]$ & 2.08 & 0.47 & 1.08 & 0.74 & 2.15 \\
$[$0.45, 0.50$]$ & 3.43 & 0.53 & 1.69 & 0.93 & 3.33 \\
$[$0.50, 0.60$]$ & 6.43 & 0.48 & 2.75 & 1.45 & 5.82 \\
	\hline
	\end{tabular*}
	\caption{Ratio of $\pizero$ production rate of MC simulation to data in the
	rapidity range 9.0$<$y$<$9.2. \label{table:ratiospectra_1}}
	\end{center}
\end{table*}
	
\begin{table*}[htbp]
	\begin{center}
	\begin{tabular*}{12cm}{@{\extracolsep{\fill}}|c c c c c c|}
	\hline
	$\pt$ range & {\sc dpmjet} 3.04 & {\sc qgsjet} II-03 & {\sc sibyll} 2.1 & {\sc
	epos} 1.99 & {\sc pythia} 8.145 \\
	$[$GeV$]$ & & & & & \\
	\hline
$[$0.00, 0.05$]$ & 1.07 & 1.29 & 0.54 & 0.98 & 1.04 \\
$[$0.05, 0.10$]$ & 1.22 & 1.33 & 0.60 & 1.10 & 1.23 \\
$[$0.10, 0.15$]$ & 1.02 & 0.95 & 0.48 & 0.89 & 1.07 \\
$[$0.15, 0.20$]$ & 0.97 & 0.78 & 0.48 & 0.77 & 1.07 \\
$[$0.20, 0.25$]$ & 1.10 & 0.70 & 0.55 & 0.86 & 1.24 \\
$[$0.25, 0.30$]$ & 1.12 & 0.52 & 0.56 & 0.78 & 1.21 \\
$[$0.30, 0.35$]$ & 1.50 & 0.47 & 0.70 & 0.66 & 1.48 \\
$[$0.35, 0.40$]$ & 2.20 & 0.41 & 0.83 & 0.66 & 1.93 \\
$[$0.40, 0.50$]$ & 6.76 & 0.59 & 1.99 & 1.46 & 5.37 \\
$[$0.50, 0.60$]$ & 15.90 & 0.26 & 3.34 & 3.12 & 11.36 \\
	\hline
	\end{tabular*}
	\caption{Ratio of $\pizero$ production rate of MC simulation to data in the
	rapidity range 9.2$<$y$<$9.4. \label{table:ratiospectra_2}}
	\end{center}
\end{table*}
	
\begin{table*}[htbp]
	\begin{center}
	\begin{tabular*}{12cm}{@{\extracolsep{\fill}}|c c c c c c|}
	\hline
	$\pt$ range & {\sc dpmjet} 3.04 & {\sc qgsjet} II-03 & {\sc sibyll} 2.1 & {\sc
	epos} 1.99 & {\sc pythia} 8.145 \\
	$[$GeV$]$ & & & & & \\
	\hline
$[$0.00, 0.05$]$ & 1.11 & 1.28 & 0.51 & 0.96 & 1.14 \\
$[$0.05, 0.10$]$ & 0.97 & 1.00 & 0.44 & 0.84 & 1.00 \\
$[$0.10, 0.15$]$ & 1.00 & 0.89 & 0.46 & 0.77 & 1.07 \\
$[$0.15, 0.20$]$ & 1.02 & 0.71 & 0.46 & 0.76 & 1.14 \\
$[$0.20, 0.25$]$ & 1.27 & 0.63 & 0.55 & 0.84 & 1.28 \\
$[$0.25, 0.30$]$ & 1.82 & 0.54 & 0.66 & 0.72 & 1.60 \\
$[$0.30, 0.35$]$ & 4.71 & 0.74 & 1.32 & 1.09 & 3.68 \\
$[$0.35, 0.40$]$ & 6.60 & 0.48 & 1.39 & 1.28 & 4.79 \\
	\hline
	\end{tabular*}
	\caption{Ratio of $\pizero$ production rate of MC simulation to data in the
	rapidity range 9.4$<$y$<$9.6. \label{table:ratiospectra_3}}
	\end{center}
\end{table*}
	
\begin{table*}[htbp]
	\begin{center}
	\begin{tabular*}{12cm}{@{\extracolsep{\fill}}|c c c c c c|}
	\hline
	$\pt$ range & {\sc dpmjet} 3.04 & {\sc qgsjet} II-03 & {\sc sibyll} 2.1 & {\sc
	epos} 1.99 & {\sc pythia} 8.145 \\
	$[$GeV$]$ & & & & & \\
	\hline
$[$0.00, 0.05$]$ & 0.98 & 1.07 & 0.39 & 0.75 & 1.00 \\
$[$0.05, 0.10$]$ & 1.04 & 0.96 & 0.42 & 0.79 & 1.09 \\
$[$0.10, 0.15$]$ & 1.23 & 0.86 & 0.49 & 0.88 & 1.24 \\
$[$0.15, 0.20$]$ & 1.66 & 0.73 & 0.56 & 0.87 & 1.45 \\
$[$0.20, 0.25$]$ & 2.96 & 0.67 & 0.71 & 0.87 & 2.23 \\
$[$0.25, 0.30$]$ & 6.42 & 0.72 & 1.07 & 1.26 & 4.30 \\
$[$0.30, 0.40$]$ & 14.86 & 0.65 & 1.51 & 2.27 & 8.85 \\
	\hline
	\end{tabular*}
	\caption{Ratio of $\pizero$ production rate of MC simulation to data in the
	rapidity range 9.6$<$y$<$10.0. \label{table:ratiospectra_4}}
	\end{center}
\end{table*}
	
\begin{table*}[htbp]
	\begin{center}
	\begin{tabular*}{12cm}{@{\extracolsep{\fill}}|c c c c c c|}
	\hline
	$\pt$ range & {\sc dpmjet} 3.04 & {\sc qgsjet} II-03 & {\sc sibyll} 2.1 & {\sc
	epos} 1.99 & {\sc pythia} 8.145 \\
	$[$GeV$]$ & & & & & \\
	\hline
$[$0.00, 0.05$]$ & 2.12 & 1.15 & 0.52 & 1.03 & 1.66 \\
$[$0.05, 0.10$]$ & 2.46 & 1.00 & 0.54 & 0.98 & 1.79 \\
$[$0.10, 0.15$]$ & 4.25 & 1.05 & 0.67 & 1.09 & 2.80 \\
$[$0.15, 0.20$]$ & 22.24 & 2.47 & 2.22 & 3.41 & 13.42 \\
	\hline
	\end{tabular*}
	\caption{Ratio of $\pizero$ production rate of MC simulation to data in the
	rapidity range 10.0$<$y$<$11.0. \label{table:ratiospectra_5}}
	\end{center}	
\end{table*}

%----- Bibliography -----

\end{document}